\documentclass{emulateapj}

\newcommand{\grb}     {GRB\,$050509$B}
\newcommand{\chan}    {{\it Chandra}}
\newcommand{\swift}   {{\it Swift}}

\newcommand{\etal}    {{\it et al.~}} 
\newcommand{\clusteralt} {{NSC~J123610$+$285901}} 
\newcommand{\cluster} {{ZwCl~1234.0$+$02916}} 
\newcommand{\galaxy} {{2MASX J12361286$+$2858580}} 
\newcommand{\dm} {{dark matter}}

\shorttitle{Host Clusters of Short GRBs}
\slugcomment{ApJ in prep: draft date \today}
\shortauthors{Dahle, H., \etal}

\begin{document}

\title{The Burst Cluster: Dark Matter in a Cluster Merger Associated with the Short Gamma Ray Burst, \grb}
\author{H.~Dahle\altaffilmark{1}, C.~L.~Sarazin\altaffilmark{2}, L.~A.~Lopez\altaffilmark{3}, C.~Kouveliotou\altaffilmark{4}, S.~K.~Patel\altaffilmark{5}, E.~Rol\altaffilmark{6}, A.~J.~van~der~Horst\altaffilmark{6}, J. Fynbo\altaffilmark{7}, R.~A.~M.~J.~Wijers\altaffilmark{6}, D.~N.~Burrows\altaffilmark{8},  N.~Gehrels\altaffilmark{9},  D.~Grupe\altaffilmark{8},  E. Ramirez-Ruiz\altaffilmark{10}, M.~J.~Micha{\l}owski\altaffilmark{7,11}} 

\email{hdahle@astro.uio.no} 

\altaffiltext{1}{Institute of Theoretical Astrophysics, University of Oslo, P. O. Box 1029, Blindern, N-0315 Oslo, Norway} 
\altaffiltext{2}{Department of Astronomy, University of Virginia, P. O. Box 400325, Charlottesville, VA 22904-4325, USA}
\altaffiltext{3}{MIT-Kavli Institute for Astrophysics and Space Research, 77 Massachusetts Avenue, 37-664H, Cambridge MA 02139, USA}
\altaffiltext{4}{Space Science Office, ZP12, NASA/Marshall Space Flight Center, Huntsville, AL 35812, USA}
\altaffiltext{5}{Optical Sciences Corporation, 6767 Old Madison Pike. Suite 650, Huntsville, AL  35806}
\altaffiltext{6}{Astronomical Institute `Anton Pannekoek', University of Amsterdam, Kruislaan 403, 1098 SJ Amsterdam, The Netherlands}
\altaffiltext{7}{Dark Cosmology Centre, Niels Bohr Institute, University of Copenhagen, Juliane Maries vej 30, DK-2100 Copenhagen, Denmark}
\altaffiltext{8}{Department of Astronomy and Astrophysics, Pennsylvania State University, 525 Davey Laboratory, University Park, PA 16802} 
\altaffiltext{9}{NASA/Goddard Space Flight Center, Greenbelt, MD 20771, USA}
\altaffiltext{10}{Department of Astronomy and Astrophysics, University of California Santa Cruz, 1156 High Street, Santa Cruz, CA 95060, USA}
\altaffiltext{11}{Scottish Universities Physics Alliance, Institute for Astronomy, University of Edinburgh, Royal Observatory, Edinburgh, EH9 3HJ, UK}

\begin{abstract}

We have identified a merging galaxy cluster with evidence of two distinct sub-clusters. The X-ray and optical data suggest that the subclusters are presently moving away from each other after closest approach. This cluster merger was discovered from observations of the first well localized short-duration gamma-ray burst (GRB), \grb. The \swift/Burst Alert Telescope (BAT) error position of the source is coincident with a cluster of galaxies \cluster, while the subsequent \swift/X-Ray Telescope (XRT) localization of the X-ray afterglow found the GRB coincident with \galaxy, a giant red elliptical galaxy in the cluster. Deep multi-epoch optical images were obtained in this field to constrain the evolution of the GRB afterglow, including a total of 27,480s exposure in the F814W band with {\it Hubble} Space Telescope Advanced Camera for Surveys (ACS), among the deepest imaging ever obtained towards a known galaxy cluster in a single passband.  We perform a weak gravitational lensing analysis based on these data, including mapping of the total mass distribution of the merger system with high spatial resolution.  When combined with \chan X-ray Observatory Advanced CCD Imaging Spectrometer (ACIS) and \swift/XRT observations, we are able to investigate the dynamical state of the merger to better understand the nature of the dark matter component. Our weak gravitational lensing measurements reveal a separation of the X-ray centroid of the western subcluster from the center of the mass and galaxy light distributions, which is somewhat similar to that of the famous ``Bullet cluster,'' and we conclude that this ``Burst cluster'' adds another candidate to the previously-known merger systems for determining the nature of dark matter, as well as for studying the environment of a short GRB. Finally, we discuss potential connections between the cluster dynamical state and/or matter composition, and compact object mergers, which is currently the leading model for the origin of short GRBs.  We also present here our results from a weak lensing survey based on archival Very Large Telescope (VLT) images in the areas of five other short GRBs, which do not provide any firm detections of mass concentrations representative of  rich clusters.
\end{abstract}


\section{Introduction} \label{sec:intro}

Dark matter comprises over 80\% of the total mass in the Universe, yet its nature is still unknown. Its presence is thus far inferred by its gravitational effects on galaxies and clusters of galaxies, and in particular cluster mergers where the galaxies and dark matter decouple from the hot gas component due to the effects of gas ram pressure as well as the weakness of the dark matter $-$ dark matter and dark matter $-$ baryon interactions. This decoupling was first clearly seen in the Bullet cluster, 1E~$0657-558$, where the two cluster cores passed through each other $\sim 100$ Myr ago \citep{Clo04,Clo06}. Weak-lensing mass reconstructions of the Bullet cluster have shown that the peaks of the mass distributions of the main cluster and the ``bullet'' sub-cluster coincide spatially with their corresponding galaxy concentrations and are offset from their X-ray halos at high significance. In 2007, another cluster merger, MACS J$0025.4+1222$, was found with clear separation of its components, enhancing the case for further identification of the presence of \dm\ in clusters \citep{Bradac08}. \cite{okabe07} analyzed seven nearby clusters at different merging stages using weak lensing observations with Subaru together with archival {\it Chandra} and {\it XMM-Newton} data. Multi-wavelength studies of one of these clusters, Abell~520, revealed a massive dark core coincident with the X-ray emission peak without a corresponding bright cluster galaxy, an observation not yet entirely understood \citep{mahdavi07}. In another cluster, A~2163, a bimodal mass distribution is observed suggesting ram pressure stripping effects \citep{okabe11,jee12} and contradictory results were presented by \cite{clowe12}. One of the most actively merging clusters, A~2744, exhibits a very complex and challenging phenomenology with a host of substructures (dark, ghost, bullet, and stripped; \citealt{merten11}). Recently, \cite{ragozzine12} reported the detection of two separate mergers taking place at the same time in the cluster A~1758, and \cite{menanteau12} reported multi-wavelength observations of  ACT$-$CL~J$0102-4915$, the most massive and most X-ray luminous cluster known at beyond redshift 0.6. Finally, analytic estimates of the time-since-collision for  DLSCL~J$0916.2+2951$ have revealed a merger at a highly evolved stage after merging ($0.7\pm0.2$\,Gyrs; \citealt{dawson12}).

Given the lack of knowledge about the 3-dimensional geometry of the mergers except for the Bullet cluster, a very simplified picture is that they contain two significant sub-clusters and a merger that occurs in the plane of the sky. This geometry results in a maximum angular offset between the hot gas and dark matter, making the physical interpretation of the system relatively straightforward. In contrast, \cite{jee07} reported the discovery of a ring of \dm\ around another cluster merger, CL~$0024+17$, where the \dm\ distribution differs from the distribution of both the galaxies and the hot gas. In this case, the merger is believed to have occurred almost exactly along the line of sight towards the cluster. Cluster mergers also constitute powerful cosmic laboratories for the study of the particle physics nature of dark matter by providing constraints on the collisional cross-section for dark matter self-interaction \citep{Mar04,Ran08} and on any high-energy emission from radiative decay of candidate dark matter particles \citep{Boy08,Rie06,Rie07a,Rie07b}. Finally, when studying the mergers of massive clusters, we witness the formation of the largest, well-defined, and virialized structures that will ever form in our Universe if the energy density is indeed dominated by dark energy. 

In this paper, we report the discovery of a merging cluster of galaxies along the line of sight of the short Gamma Ray Burst, GRB~$050509$B, based on \chan\ observations, coupled with a weak gravitational lensing analysis of HST and VLT data. \grb\ was discovered with the \swift/Burst Alert Telescope (BAT) \citep{Gehrels2004:Swift,Barthelmy2005:bat}. Subsequent X-ray observations with the \swift/X-ray Telescope (XRT) \citep{Burrows2005a:xrt} identified both the GRB afterglow and diffuse emission from a cluster of galaxies \cluster, also known as \clusteralt\ \citep{Gal2003}, at redshift $z=0.2214$ \citep{Zwicky1963}. An {\it a posteriori} probability of the chance coincidence of \grb\ with the host galaxy/cluster has been estimated as 0.1--1\% \citep{Gehrels2005,Pedersen2005,Bloom2006,Berger2007}. Our \chan\ observations of the GRB, reported here, coupled with a weak gravitational lensing analyses of VLT and HST data, show that the cluster is undergoing a merger, with indications that two distinct sub-clusters are likely moving away from each other after their closest approach.

We report the X-ray, optical, and radio observations and data analysis of \grb\ and the cluster \cluster\ in \S~\ref{sec:observ}. The results of our joint \chan\ and \swift\ spectral and spatial analysis of the X-ray emitting gas from the merging cluster and a discussion on the cluster X-ray morphology are presented in \S~\ref{sec:Xrayanalysis}. In \S~\ref{sec:lensing}, we derive the dimensionless surface mass-density distribution and the total cluster mass using observations of the burst field with the Very Large Telescope ({\it VLT}) and the Hubble Space Telescope/Advanced Camera for Surveys ({\it HST}/ACS) to measure the weak shear due to gravitational lensing. In \S~\ref{sec:WLotherGRB}, we compute weak lensing maps using public {\it VLT}\ data towards other well localized short GRBs to determine the presence (if any) of cluster-size \dm\ concentrations in their fields. In \S~\ref{sec:discuss} we discuss the possible scenarios that could account for the observed cluster structure and the putative association of \grb\ with this environment. Finally, in \S~\ref{sec:conclusion} we summarize our results. Throughout this work, we assume a cosmology with $\Omega_m = 0.3$, $\Omega_{\Lambda} = 0.7$, and $H_0 = 70$ km s$^{-1}$.    
    
\section{Observations \& Data Reduction}
\label{sec:observ}

We present below the analyses of the multi-wavelength observations of the field of \grb, the first ever short GRB, which was accurately ($<10^{\prime\prime}$) and rapidly (minutes) located \citep{Gehrels2005}. As such, the event generated a large amount of interest resulting in multiple follow-up observations (in X-ray, optical, and radio wavelengths). The study of all these data sets is invaluable for understanding both the prompt emission mechanism of the burst, the properties of its afterglow emission, its bolometric energetics, as well as the properties of its putative host galaxy and its environment. In a serendipitous way, the combination of all these wavelengths has also allowed the discovery and study of the merger results presented here. We proceed in this section to discuss the various data sets and our analyses techniques as related to the understanding of the GRB phenomenon alone.

\subsection{\swift/XRT X-ray Observations}
\label{sec:observ_swift}

XRT observed the field of \grb\ intermittently starting 2005 May 9.17 for a total of 34.5 ks of useful X-ray imaging and spectral data in Photon Counting mode. The data were processed using standard \swift\ analysis tools in the HEASOFT (v6.0.4) package and the XRT CALDB (v20060104). The \cluster\ X-ray emission is clearly detected and imaged on the center of the detector. One X-ray point source was identified and removed from this region; this source was also detected in the \chan\ observation and is discussed in \S~\ref{sec:astrometry}. The total (cluster plus background) XRT count rate is 0.0158 count/s ($0.3-10.0$ keV) and the average background, scaled to the same area and in the same energy band, is 0.0059 count/s. The XRT and \chan\ cluster spectra are discussed in detail in \S~\ref{sec:spectra}.

\subsection{\chan\ X-ray Observations} 
\label{sec:observ_chandra}

\chan\ observed the XRT error box of \grb, including the cluster, starting 2005 May 11.167 for a total useful time of 49.3 ks with the ACIS-S3 CCD in Very Faint (VF) Timed Exposure mode (ObsID 5588). The X-ray data analysis software package CIAO\footnote{http://asc.harvard.edu/ciao/} v4.1 and \chan\ CALDB\footnote{http://asc.harvard.edu/caldb/} v4.2.0 were used to process the ACIS data. We used the Repro III \chan\ data release\footnote{http://cxc.harvard.edu/ciao/repro\_iii.html}, which includes removal of background flares from the data. We then ran {\sl acis\_process\_events} to ensure that the latest gains are applied and removed the pixel randomization. Since the data were collected in VF mode, which records a $5 \times 5$ pixel region per event, we were able to do additional cleaning and removal of particle background\footnote{http://cxc.harvard.edu/cal/Acis/Cal\_prods/vfbkgrnd/}. We determined and removed the background and identified all point-like sources using {\sl wavdetect} in the \chan\ image ($0.5-8.0$~keV); a small region around each of the sources was excluded from the cluster region when extracting spectra. To account for instrumental variations, we produced exposure-corrected images. To derive a clean image of the cluster's diffuse emission (section \S~\ref{sec:wavelet}), we subsequently filled the point-source gaps with pixel count values selected from the Poisson distribution of the area surrounding the point sources using the CIAO tool {\sl dmfilth}.

\subsection{VLT Optical data} \label{sec:observ_VLT}  
We downloaded from the {\it VLT} archive all public data taken with the FOcal Reducer and low dispersion Spectrographs 1 and 2 (FORS1 and FORS2) for \grb\ and five other well-localized short GRBs. FORS1 has a pixel size of 0\farcs 2 and a field of view of $6\farcm 8 \times 6\farcm 8$, while the pixels of FORS2 were re-binned to a pixel scale 0\farcs 25, with a usable field of view of $6\farcm 9 \times 6\farcm 9$. The data were reduced in a standard fashion using IRAF\footnote{IRAF is distributed by the National Optical Astronomy Observatories, which are operated by the Association of Universities for Research in Astronomy, Inc., under a cooperative agreement with the National Science Foundation.}; overscan correction and  bias subtraction were performed separately on the four read-out quadrants of the FORS1 chips. The reduced images were corrected for cosmic rays using the Laplacian cosmic-ray identification method of \cite{vanDokkum2001}. Finally, individual images were interpolated to the same reference frame using a third order polynomial and were added together. Images were not added across intervals spanning more than several hours, since in most cases, circumstances such as the seeing were significantly different between the various epochs.

As shown in Table~\ref{tab:VLTdata}, the number of galaxies detected in the images of a given GRB field varied significantly between different epochs, due to variations in exposure time, sky brightness and seeing. The Galactic extinction was also significantly different across the various GRB fields. 

Finally, for \grb\ only, we aligned the {\it VLT} image to the 2MASS coordinate system as the latter is used in  \S~\ref{sec:astrometry} to register the X-ray images. To this end, we searched the central $6\farcm5 \times 3\farcm6$ region of the 3600 s $R-$band {\it VLT}/FORS2 image using LEXTRCT (with $\sigma_{\rm psf}=2.216$ pixels and SN ratio$=1.1$) and detected 1515 sources. We then searched the 2MASS catalog and found 14 acceptable counterparts to our $R-$band sources, which we used to align the {\it VLT} image to the 2MASS coordinate system to an accuracy of $\sigma_{\rm rms}({\rm RA})=0\farcs17$ and $\sigma_{\rm rms}({\rm DEC})=0\farcs21$.

\begin{deluxetable*}{lccccrcr} 
\tablecolumns{8} \tablewidth{0pt}
\tablecaption{Public {\it VLT} Data for six Well-Localized Short GRBs \label{tab:VLTdata}}
\tablehead{ \colhead{GRB} & \colhead{Galactic extinction} &  \colhead{Date} &  \colhead{Instrument} &
\colhead{Filter} & \colhead{$t_{\rm exp}$} & \colhead{FWHM} &  \colhead{$N_{\rm gal}$} \\
\colhead{name} & \colhead{$E(B-V)$} & \colhead{(UT)} & \colhead{name} & \colhead{ } & \colhead{(s)} & \colhead{(arcsec)} & \colhead{ }  
}
\startdata
           \objectname{050509B}  & 0.019 & 2005 May 13.07 & FORS2 &  $V$ & 2700 & 0.63 & 2674 \\
             & & 2005 Jun 01.02 & FORS2 &  $V$ & 1800 & 0.81 & 1883 \\
             & & 2005 May 11.00 & FORS2 &  $R$ & 2160 & 0.85 & 1651 \\
             & & 2005 May 31.96 & FORS2 &  $R$ & 3600 & 0.70 & 2257 \\
           \objectname{050724} & 0.590 & 2005 Jul 24.99 & FORS1 &  $V$ & 480 & 0.82 & 543 \\
                        & & 2005 Jul 27.99 & FORS1 &  $V$ & 540 & 0.77 & 556 \\
                        & & 2005 Jul 25.00 & FORS1 &  $R$ & 540 & 1.08 & 277 \\
                        & & 2005 Jul 25.98 & FORS1 &  $R$ & 540 & 0.99 & 417 \\
                        & & 2005 Jul 27.97 & FORS1 &  $R$ & 540 & 0.77 & 537 \\	                
		                    & & 2005 Jul 30.10 & FORS1 &  $R$ & 720 & 0.48 & 672 \\	                	
                        & & 2005 Aug 22.99 & FORS1 &  $R$ & 780 & 1.45 & 359 \\
                        & & 2005 Aug 25.98 & FORS1 &  $R$ & 1140 & 0.68 & 815 \\
                        & & 2005 Jul 25.01 & FORS1 &  $I$ & 540 & 0.95 & 483 \\
                        & & 2005 Jul 25.97 & FORS1 &  $I$ & 360 & 0.77 & 483 \\
                        & & 2005 Jul 27.98 & FORS1 &  $I$ & 540 & 0.74 & 496 \\
                        & & 2005 Jul 30.11 & FORS1 &  $I$ & 720 & 0.50 & 733 \\                        
           \objectname{050813}  & 0.054 & 2005 Aug 19.06 & FORS2 & $I$ & 1800 & 0.64 & 1048 \\
           \objectname{050906}  & 0.066 & 2005 Sep 25.28 & FORS1 & $V$ & 1800 & 0.89 & 1541 \\
                        & & 2005 Sep 07.35 & FORS2 &  $V$ & 1800 & 0.65 & 2248 \\
                        & & 2005 Sep 12.31 & FORS2 &  $V$ & 1800 & 0.91 & 1611 \\
                        & & 2005 Sep 25.27 & FORS1 &  $R$ & 1800 & 0.88 & 2308 \\   
                        & & 2005 Sep 07.33 & FORS2 &  $R$ & 1800 & 0.70 & 2088 \\
                        & & 2005 Sep 12.30 & FORS2 &  $R$ & 1800 & 0.79 & 2050 \\
                        & & 2005 Sep 25.29 & FORS1 &  $I$ & 1800 & 0.82 & 1759 \\	                 
                        & & 2005 Sep 07.37 & FORS2 &  $I$ & 1920 & 0.73 & 1497 \\
                        & & 2005 Sep 12.32 & FORS2 &  $I$ & 1440 & 0.86 & 1410 \\
           \objectname{050911} 
                        & 0.010 & 2005 Sep 12.19 & FORS1 &  $V$ & 540 & 0.57 & 1512 \\
                        & & 2005 Sep 12.41 & FORS1 &  $V$ & 540 & 0.77 & 1359 \\
                        & & 2005 Sep 12.18 & FORS1 &  $R$ & 540 & 0.56 & 1727 \\  
                        & & 2005 Sep 12.40 & FORS1 &  $R$ & 540 & 0.74 & 1061 \\
           \objectname{060313} & 0.064 & 2006 Mar 20-30 & FORS1 &  $R$ & 9190 & 1.21 & 2158 \\
\enddata
\end{deluxetable*}

\subsection{HST/ACS Optical Data} \label{sec:observ_ACS}

The field of \grb\ was observed for a total of 27,480 s, using the Wide Field Channel (WFC) of the {\it HST}/ACS. The imaging data were obtained with the F814W filter on four different dates in 2005 May and June (Proposal ID 10119). Data were retrieved from the {\it HST} archive in the form of ACS association FITS images (i.e., each image was based on 3 individual dithered exposures), which had been calibrated using the {\it MultiDrizzle} pipeline software. This calibration pipeline produces cosmic-ray cleaned images of pixel scale 0\farcs 05, which have been re-sampled and combined using the drizzle software of \cite{Fruchter2002}, while correcting for the geometric distortion of ACS.

The resulting set of 12 images were offset by up to $2\arcsec$ in $x$ and $y$ (and also slightly rotated with respect to each other), and needed re-registration before combination into a single, deep image. This was done by running the IMCAT \citep{Kaiser1995} peak finding algorithm ({\it hfindpeaks}) on the FITS image corresponding to each observation set, and performing aperture photometry ({\it apphot}) on the detected objects.  Approximately 200 stars and bright, compact non-stellar objects were selected as reference objects for the re-registration of the images. A fiducial frame was chosen among the available images, and the transformation of each of the other images into this frame was calculated as a second-order polynomial. The solutions for these transformations were obtained iteratively, by interactive rejection of outliers to the fit until a good solution was found with rms residuals less than 0.1 pixel in $x$ and $y$. Based on the resulting transformation parameters, all images were then re-registered into the same coordinate system and combined into a single, deep frame. The image is comparable in depth to the combined F814W frames of the original Hubble Deep Fields \citep{Wil96}, and is one of the deepest images ever obtained towards a galaxy cluster beyond Coma in a single passband.

\subsection{Radio Observations} \label{sec:observ_radio}

\grb\ was observed at centimeter wavelengths with the Westerbork Synthesis Radio Telescope (WSRT). We used the Multi Frequency Front Ends \citep{Tan1991} in combination with the IVC$+$DZB backend\footnote{Section 5.2 at http://www.astron.nl/wsrt/wsrtGuide/node6.html} in continuum mode, with a bandwidth of $8 \times 20$ MHz. The observations were carried out on two epochs in 2005 (Table~\ref{tab:radio}). On the first epoch, the source was observed for 12 hours at 6 cm (4.86 GHz); on the second epoch we switched between 6 cm and 21 cm (1.43 GHz), with 40 minutes dwell time per frequency. Gain and phase calibrations were performed with the calibrator 3C~286. Data reduction was performed using the MIRIAD software \citep{Sault1995}. In Table~\ref{tab:radio} we report the upper limits on the 4.9 and 1.4 GHz fluxes at the location of the \swift/XRT GRB afterglow. Preliminary results of the afterglow flux from the first epoch were published in \citet{vanderHorst2005}.

We detected a radio source, G1, in the center of the GRB host galaxy. The source is unresolved at the WSRT resolution, with a synthesized beam size (full width at half maximum) of $24\farcs7 \times 9\farcs5$ at 1.4~GHz and $17\farcs6 \times 8\farcs3$. in the combined image at 4.9~GHz. Using the flux of the combined dataset at 4.9 GHz, we derived a spectral index of $-1.07 \pm 0.26$ ($1.4-4.9$ GHz). Two possible origins for the radio source are either a mildly active galactic nucleus or nuclear star formation. The latter is unlikely to be the correct explanation as our data would imply a star formation rate (SFR) of $71\pm21\;\mbox{M}_{\odot}\;\mbox{yr}^{-1}$ [following \cite{Berger2003}]. This value is much higher than the $3\;\sigma$ upper limit to the current SFR of $0.1\;\mbox{M}_{\odot}\;\mbox{yr}^{-1}$ found for the host galaxy from the H$\alpha$ luminosity \citep{Bloom2006}. Note that if the radio emission is due to star formation, much of it is expected to come from supernova remnants (SNR), and it would thus also depend on the SNR history. Finally, we do not expect to have a large fraction of obscured star formation by dust in this elliptical galaxy.
On the other hand, the value of the radio spectral index is quite typical for optically thin AGN jets, although a bit at the high end of the distribution \cite{brown91}. If the emission would be dominated by the optically thick core of the jet, the spectrum would be quite flat.

\begin{deluxetable*}{lcccc}
\tablecolumns{5} 
\tablewidth{0pt}
\tablecaption{WSRT Measurements of the \grb\ Afterglow and Host Galaxy Flux \label{tab:radio}}
\tablehead{ \colhead{Frequency} & \colhead{Observation Date} & \colhead{Time Since GRB Trigger}  & \colhead{GRB Flux\tablenotemark{a}} & \colhead{Galaxy Flux} \\
\colhead{(GHz)}       & \colhead{(UT)}               & \colhead{(days)}    & \colhead{($\mu$J)}                    & \colhead{($\mu$J)} }
\startdata
4.9  &  May 09.63-10.12            &  00.46-00.96 & $31\pm 22$ & $315\pm 61$ \\
4.9  &  May 19.60-20.08            & 10.43-10.92 & $40\pm 33$ & $325\pm 76$ \\
4.9  &  May 9 \& 19 (combined) & -- & $11\pm 20$ & $321\pm 59$ \\
1.4  &  May 19.63-20.10             & 10.46-10.93 & $1\pm 17$  & $1183\pm 195$ \\
\enddata
\tablenotetext{a}{Formal flux measurements at the GRB position with $1\sigma$ rms noise around that position.}
\end{deluxetable*}

\section{X-ray Data Analysis and Results} \label{sec:Xrayanalysis}

\subsection{Astrometry and X-ray Source Catalog}
\label{sec:astrometry}

We describe here the results of our X-ray source search in the \chan/ACIS and \swift/XRT images. We searched the \chan\ image for X-ray sources at the highest spatial resolution (0\farcs492 per pixel) and within a radius of $\sim8\arcmin$ of the nominal aimpoint. We used the point source detection method (Gaussian PSF fitting) described in \citet{Tennant2006}\footnote{Available as part of the image analysis package LEXTRCT (http://wwwastro.msfc.nasa.gov/qdp)}. Only events in the energy range $0.5-7.0$ keV were used, to maximize the source to background count ratio. The search resulted in 43 X-ray sources with a signal-to-noise (SN) ratio $> 2.8$. The sources are listed in Table~\ref{xraysourcelist}, which gives their positions (after the \chan\ image was aligned to the 2MASS catalog as described below), their net number of counts, the width of the \chan\ PSF at each source location on the ACIS-S chip, and the SN ratio of each source.

We then searched for counterparts within 2\arcsec\ of each \chan\ X-ray source in the 2MASS Point Source Catalog \citep{Skr06}. Only two counterparts were found (these are \#6 and \#12 in Table~\ref{xraysourcelist}). We used these two sources and two different alignment methods (CIAO/wavdetect and LEXTRCT) to compute a mean shift between the \chan\ and 2MASS images. The results from both methods were consistent and gave a shift of $<0\farcs4$ in each coordinate. We also compared our X-ray source list to the USNO2B catalog and found 9 potential matches.  As the errors in the USNO source positions are known to be typically greater than the uncertainties in the 2MASS positions, we elected to tie all fields to the 2MASS coordinate system throughout this paper.

\begin{deluxetable*}{lccrcr}
\tabletypesize{\footnotesize}
\tablecolumns{6} 
\tablecaption{\chan\ X-ray source list \label{xraysourcelist}}
\tablewidth{0pt}
\tablehead{ \colhead{Source ID} & \colhead{RA}    & \colhead{DEC}   & \colhead{Net Counts} & \colhead{$\sigma_{\rm psf}$} & \colhead{SN}\\
 & \colhead{(h m s)} & \colhead{(\arcdeg\ \arcmin\ \arcsec)} & &\colhead{(pix)} & ratio }
\startdata
\hline \\
  1  & 12 35 47.211  & 29  03 30.619  & 136.90  & 6.37  &10.88  \\
  2  & 12 35 48.503  & 28 58 54.896  &  15.52  & 3.62  & 3.49  \\
  3  & 12 35 50.664  & 28 57 39.498  &  47.54  & 3.23  & 6.13  \\
  4  & 12 35 52.105  & 29  05 22.097  &  22.13  & 7.49  & 4.13  \\
  5  & 12 35 52.470  & 28 57 56.633  &  45.07  & 2.81  & 6.02  \\
  6$^*$  & 12 35 52.699  & 28 58 56.507  &  17.42  & 2.73  & 3.78  \\
  7  & 12 35 55.091  & 29  05 43.247  &  48.49  & 7.50  & 6.36  \\
  8  & 12 35 55.542  & 29  00 42.569  &  22.18  & 2.66  & 4.01  \\
  9  & 12 35 56.577  & 28 59 46.196  &  18.37  & 2.19  & 3.80  \\
 10  & 12 35 57.387  & 28 56  04.520  &  58.91  & 2.63  & 6.70  \\
 11  & 12 35 58.058  & 28 57 49.011  &  14.79  & 1.92  & 3.44  \\
 12$^*$  & 12 36  01.747  & 29  01  00.472  &  24.26  & 1.98  & 4.45  \\
 13  & 12 36  02.871  & 28 58 35.169  &  42.70  & 1.30  & 5.72  \\
 14  & 12 36  03.407  & 29  00 14.984  & 104.92  & 1.52  & 9.60  \\
 15  & 12 36  03.901  & 29  01 24.909  &  57.63  & 2.01  & 6.77  \\
 16  & 12 36  05.114  & 29  02 23.551  &  24.96  & 2.58  & 4.64  \\
 17  & 12 36  05.146  & 29  00 13.265  &   9.86  & 1.37  & 2.81  \\
 18  & 12 36  06.909  & 29 00 19.693  &  14.14  & 1.31  & 3.61  \\
 19  & 12 36  07.307  & 29 00 18.046  &  22.76  & 1.28  & 4.42  \\
 20  & 12 36 07.768  & 28 58 02.824  & 110.38  & 1.02  & 9.09  \\
 21  & 12 36  08.988  & 28 58 51.684  &  13.95  & 0.94  & 3.13  \\
 22  & 12 36  09.353  & 28 56  00.946  &  77.11  & 1.66  & 7.70  \\
 23  & 12 36 10.376  & 29  02 37.228  & 127.56  & 2.55  &10.05  \\
 24  & 12 36 10.405  & 29  01  05.592  & 278.17  & 1.53  &15.23  \\
 25  & 12 36 11.964  & 28 58 47.042  &  13.79  & 0.90  & 3.09  \\
 26  & 12 36 12.481  & 29  02 21.743  & 250.54  & 2.34  &14.41  \\
 27  & 12 36 14.526  & 29  00 41.834  & 186.14  & 1.37  &12.29  \\
 28  & 12 36 16.035  & 29  02 44.268  &  18.62  & 2.73  & 3.84  \\
 29  & 12 36 19.712  & 29  04 37.591  &  38.45  & 4.92  & 5.23  \\
 30  & 12 36 20.887  & 29  03 38.672  &  15.28  & 3.90  & 3.25  \\
 31  & 12 36 21.484  & 29  01 49.972  &  21.14  & 2.42  & 3.95  \\
 32  & 12 36 22.262  & 29  03 52.589  &  26.00  & 4.29  & 4.43  \\
 33  & 12 36 23.586  & 29  01 43.497  &  46.15  & 2.57  & 6.12  \\
 34  & 12 36 23.734  & 29  02 53.964  &  11.35  & 3.49  & 2.85  \\
 35  & 12 36 23.768  & 29 00 56.249  &  11.08  & 2.15  & 3.10  \\
 36  & 12 36 23.817  & 28 59 46.703  &  54.18  & 1.75  & 6.46  \\
 37  & 12 36 29.356  & 29  01 38.599  & 394.58  & 3.36  &17.48  \\
 38  & 12 36 30.279  & 28 58 52.432  &  34.99  & 2.60  & 5.07  \\
 39  & 12 36 31.187  & 29  00 30.065  &  35.91  & 3.12  & 5.32  \\
 40  & 12 36 31.240  & 29  01 42.171  &  32.02  & 3.75  & 4.73  \\
 41  & 12 36 31.716  & 29  03  03.138  &  80.40  & 4.88  & 7.76  \\
 42  & 12 36 39.217  & 28 59 23.538  &  46.13  & 4.69  & 5.79  \\
 43  & 12 36 40.077  & 29  00 57.903  &  31.26  & 5.44  & 4.05  \\
\enddata
\tablecomments{1 pixel = 0\farcs492. Coordinates are referenced to the 2MASS coordinate system, and RA and DEC are accurate to 0.4\arcsec (see $\S$\ref{sec:astrometry}). $\sigma_{\rm psf}$ is the width of the PSF at the pixel location of the point source on the ACIS-S3 chip, as given by LEXTRCT $^*$ 2MASS source}
\end{deluxetable*}

We used the PSF fitting method described above to detect sources on the \swift/XRT image in the energy band $0.3-10$ keV. Here, we adopted a fixed PSF FWHM of 3\farcs77 (1.6 XRT pixels) based on the on-ground PSF calibration tests of the XRT using the Al K line at 1.49 keV \citep{Moretti2004}. We detected 16 sources with SN ratio $> 2.8$ (SN ratio$_{\rm min}=2.80$ resulted in 9.4 net counts). Each source was individually inspected and all appeared to be real point-like X-ray sources and not merely background fluctuations. All the XRT detected sources that fell on our \chan\ field of view were within 3 XRT pixels of the \chan\ position. We excluded one source, which was close to the detector edge and might have a less reliable position estimate, and used the remaining 7 common sources between the \chan\ and XRT images to derive a transformation from XRT to sky coordinates. The shifts in RA and DEC were $+0\farcs15$ and $-1\farcs20$, respectively. We attribute the difference between the offset reported here and the offsets reported in \citet{Burrows2005b} and \citet{Bloom2006} to the refined boresight calibration \citep{Moretti2006}, which was not applied previously to the analyzed data.

We presented the earlier results of our re-analysis for the XRT position of the \grb\ afterglow in \cite{Burrows2005b}. Our derived positions for the source centroid using PSF-fitting with LEXTRCT and with CIAO/{\it wavdetect} are consistent with the location reported by \cite{Burrows2005b} at RA(J2000) $= 12^{\rm h} 36^{\rm m} 13.58^{\rm s}$, DEC(J2000) $= 28\arcdeg 59\arcsec 01\farcs3$ (error radius of $9\farcs3$ at the 90\% confidence level).
  
\subsection{X-ray Image Analyses} \label{sec:Xray_image}

\subsubsection{$\beta$ model fits} \label{sec:beta_fits}

\begin{figure}
\includegraphics[width=\columnwidth]{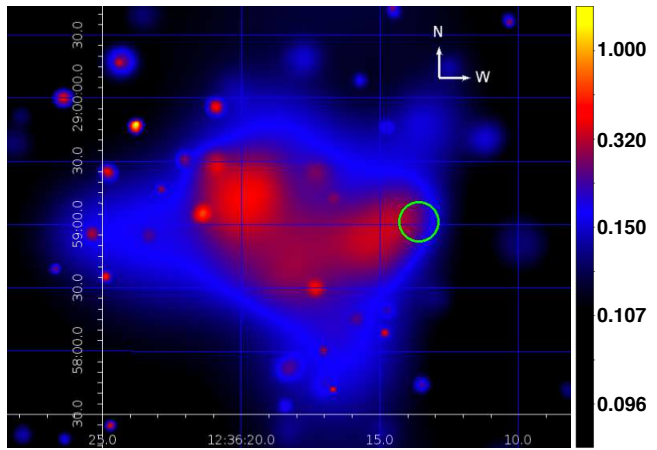}
\caption{Adaptively smoothed \chan\ X-ray image of \cluster; image is $3\farcm4 \times 3\farcm0$ in size and the color bar is normalized to one. The detected point sources have been removed prior to smoothing. The point-like features seen in the image indicate the presence of slightly extended sources, or fluctuations in the background, which were exaggerated by the smoothing and exposure corrections. The green circle ($r = 9\farcs3$) is the 90\% confidence error region of the XRT X-ray afterglow of \grb.}
\label{xrayimages}
\end{figure}

To qualitatively characterize the distribution of the diffuse X-ray emission we utilized adaptive smoothing (using CIAO {\it csmooth}) to create the image shown in Figure \ref{xrayimages}. We note that the image shows that the gas distribution in \cluster\ is bimodal, indicative of a merger between the main cluster component to the east and a subcluster component to the west. The strongest emission is associated with the eastern subcluster, and roughly centered on the galaxies 2MASX J12362010$+$2859080 and 2MASX J12362094$+$2859290. There may be a region of concentrated X-ray emission centered on the brightest cluster galaxy 2MASX J12362094$+$2859290, and a more diffuse eastern component located to the southwest of this galaxy, and roughly centered on 2MASX J12362010$+$2859080. The western subclump in the X-ray image is elongated.

The X-ray surface brightness declines somewhat rapidly to the west, while there is a long tail of emission to the east.
The brightest part of the western X-ray emission is displaced from the GRB host galaxy, \galaxy, even though this is the dominant galaxy in the western subcluster. The overall morphology of the western subclump resembles that seen in clusters with a merger ``cold front'' \citep{Mar+00,VMM01}; unfortunately, the existing \chan\ observation is not deep enough to allow either a sharp surface brightness edge or a temperature jump to be detected at a statistically significant level. The elongated tail on the east of the western subcluster, the drop in surface brightness to the west of the western peak, and the separation between the western X-ray peak and the dominant galaxy in this region, all suggest that the western subclump is moving to the west relative to the bulk of the gas in the eastern subcluster, and that the gas in the western subclump has been slowed and shaped by ram pressure. This morphology resembles that seen in numerical simulations of offset, slightly unequal mass mergers, just after first core passage
\citep[e.g., see Fig.~4 and Fig.~18b in][]{poole06}.

We proceed below employing two methods to quantitatively describe the density distribution of the X-ray emitting gas in the cluster.

\subsubsection{Wavelet-Transform Analysis} \label{sec:wavelet}

The first method we used to quantitatively describe the morphology of the diffuse X-ray emission is the wavelet-transform analysis on the full-band ($0.5-8.0$ keV) ACIS image after point-source removal. Here we followed the method outlined in \cite{lopez09a}: we convolved every pixel of the \chan\ image with Mexican-hat wavelet functions of different widths. A Mexican-hat function is the optimum wavelet to analyze X-ray images because it removes flat features, like noise, and has a shape similar to a Gaussian signal \citep{lopez09a}. Wavelet-transformed images are produced by calculating the summed intensity enclosed by the area of the Mexican hat. The resulting transformed images thus filter the original signal intensity at the scale of each wavelet, producing a sequence of images over a range of physical scales. Consequently, the technique can be used to measure the size and distribution of X-ray emitting substructures. 
 
\begin{figure*}
\begin{center}
\includegraphics[width=0.65\textwidth]{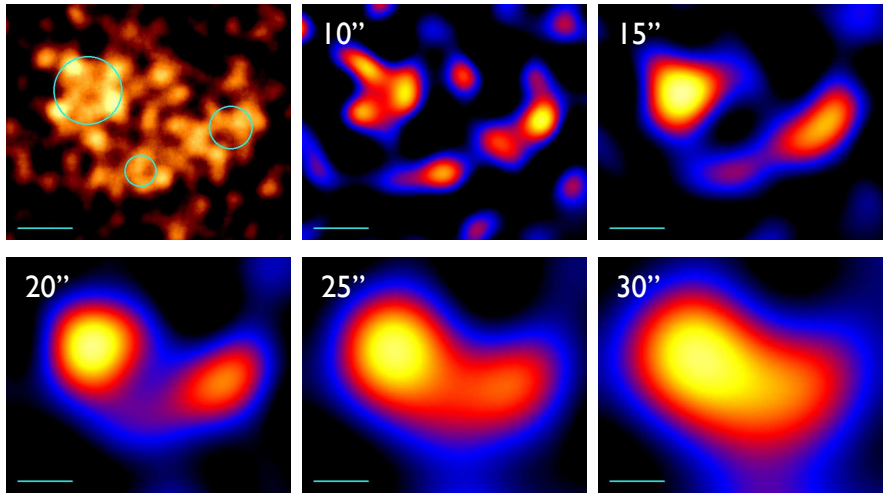}
\end{center}
\caption{The wavelet-transformed images of \cluster\ at five different scales.
Here east is to the left and north is up.
Each image shows the emission at the scale of the wavelets.
The scale bar is 100 kpc ($\approx 28\farcs5$) assuming an angular-size distance of 729.5 Mpc.
The leftmost panel is the full X-ray band image smoothed with a 5\arcsec\ Gaussian.
The three cyan circles indicate the centroids of the structures listed in Table~\ref{Xcentroids}. The following five panels are the wavelet-transformed images. At the smallest scale (10\arcsec) many structures are evident. At intermediate scales (15\arcsec and 20\arcsec), the source is characterized by two emitting regions bridged together. At the largest scales (25\arcsec and 30\arcsec), the source appears as a single emitting region.}
\label{montage}
\end{figure*}

Fig.~\ref{montage} shows the wavelet-transformed images of \cluster\ at five different scales. The 10\arcsec\ image shows numerous individual structures that have a similar distribution to the X-ray surface brightness contours shown in Fig.~\ref{centroid_combination}. At larger scales (15\arcsec\ and 20\arcsec), the source has two identifiable emitting regions that are connected by a narrow, extended ``bridge" of emission. Finally, at the largest scales, the source appears as a single emitting region. The bimodal structure evident in the transformed images supports the cluster merger interpretation of \S~\ref{sec:beta_fits}.   

We calculated the centroids and sizes of the Eastern and Western subclusters and the bridge in the X-ray image using wavelet-transform analysis. Specifically, the sizes of the structures can be characterized by the scale at which the convolution of the wavelets and the structures are maximum; the centroids of the structures are given by the pixel locations where the transformed images have the largest values \citep{lopez09a}. Table~\ref{Xcentroids} lists their derived locations (RA and DEC) in the \chan\ image and their sizes in arcseconds. We estimate the projected separation between the Eastern and Western clusters (centroid to centroid) to be 73\arcsec, corresponding to $\sim$260 kpc at the cluster distance.

\begin{deluxetable*}{lccr} 
\tablecolumns{4} \tablewidth{0pt} 
\tablecaption{Cluster X-ray Substructure Properties \label{Xcentroids}}
\tablehead{ \colhead{Substructure} & \colhead{RA} & \colhead{DEC} & \colhead{size} \\
     &    \colhead{(h m s)} & \colhead{(\arcdeg\ \arcmin\ \arcsec)}  &   \colhead{(\arcsec)}   }
\startdata
Eastern Cluster	 & 12 36 20.29 & $+$28 59 14.74 & 17.2 \\
Western Cluster	 & 12 36 14.75 & $+$28 58 55.84  & 10.8 \\
Bridge	                 & 12 36 18.25 & $+$28 58 33.81  & 7.9 \\
\enddata
\tablecomments{Derived using wavelet-transform analysis detailed in $\S$\ref{sec:wavelet}. The estimated uncertainties in the centroid values are 4 pixels $\approx2\arcsec$.}
\end{deluxetable*}

\subsubsection{Power-Ratio method} \label{sec:power_ratios}

Next, we used the power-ratio method (PRM) on the full-band ($0.5-8.0$ keV) ACIS image of the \cluster\ after point-source removal to quantify the X-ray morphology of \cluster. This method was originally developed to probe the dynamical state of clusters observed with {\it ROSAT} \citep{buote1995,buote1996} and was extended to \chan\ by \cite{jeltema05}. A detailed description of the method is given in \cite{jeltema05}; below we present a qualitative description. The PRM measures asymmetries in an image using the multipole moments and the associated powers $P_{\rm m}$ of the surface brightness in a circular aperture. The higher-order terms measure asymmetries at successively smaller scales relative to the X-ray object size. We divided all powers by $P_{\rm 0}$ to form the power ratios, $P_{\rm m}/P_{\rm 0}$, to normalize with respect to flux. Morphological information is given primarily by the higher-order terms $P_{\rm 2}/P_{\rm 0}$ (quadrupole ratio) and $P_{\rm 3}/P_{\rm 0}$ (octupole ratio), which are sensitive to the source ellipticity and to deviations from mirror symmetry, respectively.  Sources with high $P_{\rm 2}/P_{\rm 0}$ have elliptical or elongated morphologies; high $P_{\rm 3}/P_{\rm 0}$ ratios occur in sources with
mirror asymmetry, such as bimodal structures of unequal sizes.

\begin{figure} 
\includegraphics[width=0.95\columnwidth]{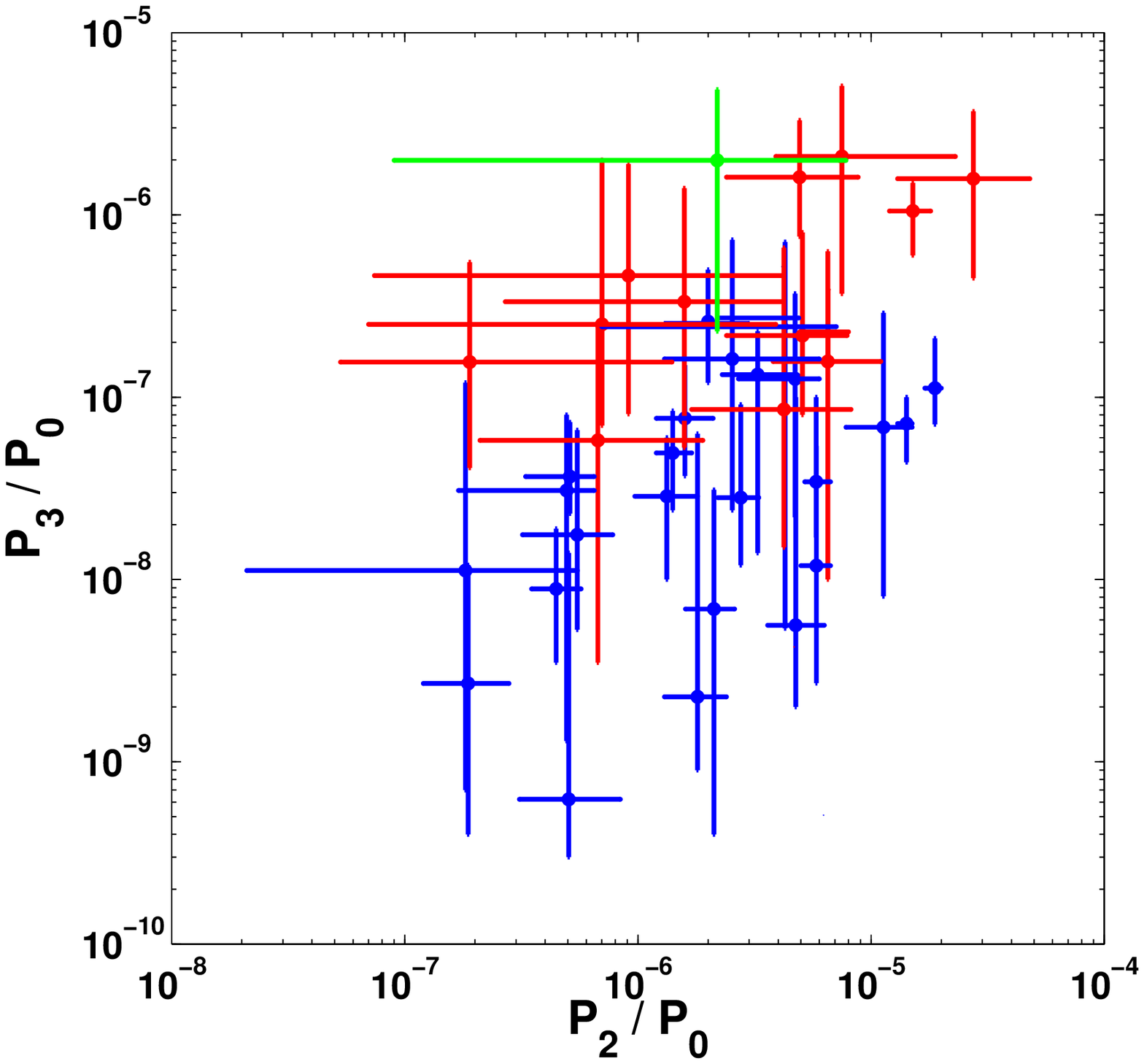}
\caption{The $P_{\rm 3}/P_{\rm 0}$ {\it versus} $P_{\rm 2}/P_{\rm 0}$ of the clusters in the \cite{jeltema05} paper with the value of \cluster\ added in green. The red (blue) points are clusters with $z>0.5$ $(<0.5)$, respectively.}
\label{powers}
\end{figure}

Fig.~\ref{powers} plots our derived $P_{\rm 3}/P_{\rm 0}$ {\it versus} $P_{\rm 2}/P_{\rm 0}$ using an aperture of 0.5 Mpc with the values for the clusters discussed in \cite{jeltema05}; the ratio values are $P_{\rm 3}/P_{\rm 0}=(19.9^{+28.8}_{-17.6}) \times 10^{-7}$ and $P_{\rm 2}/P_{\rm 0}=(21.9^{+56.2}_{-21.0}) \times 10^{-7}$. The errors on the plot were estimated using the \cite{jeltema05} technique and are at the 90\% confidence level. We note that the $P_{\rm 3}/P_{\rm 0}$ value of \cluster\ is the second largest in that group and a factor of $\sim30$ higher than the average for $z<0.5$, indicating a highly asymmetric (bimodal) structure. These results suggest that the merger in  \cluster\ has occurred recently, as the cluster is quite asymmetric and is not yet relaxed \citep{jeltema05}. For example, the simulations shown in \citet[][Fig.~13]{poole06} show that such large values primarily occur before second core passage.

\subsection{Cluster X-ray Spectral Analysis} \label{sec:spectra}

We extracted a \swift/XRT source spectrum from a region 1\farcm7 in radius centered on the cluster emission (at 12$^{\rm h}$36$^{\rm m}$18.848$^{\rm s}$, $+$28$^{\circ}$59$\arcmin$12.17$\arcsec$, J2000.0) and created the background spectrum using events from a $2\arcmin-7\arcmin$ annulus about the cluster center; all point sources were excluded from the source and the background regions. We restricted all XRT imaging and spectral analysis to events with energies between $0.3-10.0$ keV and modeled the cluster spectrum using a MEKAL model for X-ray emission from an optically thin plasma \citep{Mewe1985}. To justify use of the $\chi^2$ statistic, we grouped spectral bins to obtain at least 10 counts per bin before background subtraction. We also accounted for absorption due to intervening gas and dust from the Galaxy using the Tuebingen-Boulder ISM absorption model (XSPEC model {\sl tbabs}) with the relative abundances and cross-sections from \citet{Wilms2000} and with the He cross section from \citet{Yan1998}; the inclusion of a redshifted absorption component (XSPEC model {\sl ztbabs}) did not result in an improved fit and was not used further in the spectral analysis. We initially held the column density constant at the average Galactic value toward the cluster derived from radio measurements, $N_{\rm H}=1.52\times10^{20}$~cm$^{-2}$ \citep{Dickey1990}, and the metallicity at the mean value, $Z=0.41$~solar, as derived from spectral fits to 38 clusters \citep{Bonamente2006}. The best fit spectral model parameters are reported in Table \ref{globalspec}.  When we let the cluster metallicity vary, we found no statistically significant improvement in the fit and we were only able to place a 90$\%$ upper limit of $Z<0.9$~solar. For the remainder of the XRT analysis we, therefore, fixed this parameter to the fiducial value of 0.41 solar. 

\begin{figure}
\includegraphics[angle=-90,width=0.95\columnwidth]{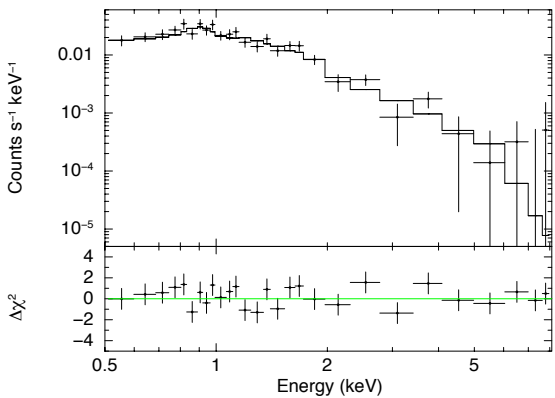}
\caption{\chan/ACIS-S spectrum of the cluster \cluster. The best fit absorbed MEKAL model and residuals are also plotted.}
\label{fig:globalspectrum}
\end{figure}

We collected a \chan/ACIS source spectrum from a circular region 1\farcm7 in radius centered on the cluster, and extracted a background spectrum from an annular region $1\farcm7-2\farcm8$ (1.5 times the source area) also centered on the cluster; all point sources were excluded from the source and the background regions. We generated a position-weighted ancillary response using the CIAO script {\sl specextract} for each spectrum; while running this script, the energy redistribution matrix is generated using {\sl mkacisrmf}. We analyzed the data by using the XSPEC (v12.3.1) spectral-fitting package \citep{Arnaud1996}. To justify use of the $\chi^2$ statistic, we grouped spectral bins to obtain at least 10 counts per bin before background subtraction. We fit the $0.5-8.0$ keV energy band with the same absorbed MEKAL model that was applied to the XRT data. We first fit with $N_{\rm H}$ and metallicity fixed and obtained the fit described in Table \ref{globalspec}. We then freed only the metallicity, which resulted in a best fit abundance of $0.3^{+0.5}_{-0.3}$ solar.  Next, we restored the metallicity to 0.41 solar and freed the column density, which resulted in $N_{\rm H}=1.0^{+4.5}_{-1.0} \times10^{20}$~cm$^{-2}$; in both cases, however, the change in $\chi^2$ was not significant, and the fit parameters were, therefore, not listed in Table \ref{globalspec}. Figure \ref{fig:globalspectrum} shows the ACIS spectrum.

Finally, we measured the temperatures/luminosities of the two X-ray subclusters over the $0.5-10$ keV range and we find for the Eastern subcluster: $kT=3.2^{+1.6}_{-0.9}$, Unabs $F_{\rm x}=0.53^{+0.12}_{-0.09} \times 10^{-13}$ erg s$^{-1}$ cm$^{-2}$, $L_{\rm x} = 7.8^{+1.7}_{-1.3} \times 10^{42}$ erg s$^{-1}$, $\chi^2$/dof=67/72, and for the Western subcluster $kT=3.3^{+3.7}_{-1.2}$, Unabs $F_{\rm x}=0.17\pm0.02 \times 10^{-13}$ erg s$^{-1}$ cm$^{-2}$, $L_{\rm X} = 2.5\pm0.3 \times 10^{42}$ erg s$^{-1}$, $\chi^2$/dof=23/25. We note that their summed flux does not equal the total X-ray luminosity, indicating that there is non-negligible flux outside the two subclusters.

\begin{deluxetable*}{lcccccc}  
\tablecolumns{7} \tabletypesize{\footnotesize}
\tablecaption{Global X-ray Cluster Spectral Properties \label{globalspec}}
\tablewidth{0pt}
\tablehead{ \colhead{} & \colhead{$N_{\rm H}$}  & \colhead{$kT^{\rm a}$}   & \colhead{$Z$}      & \colhead{Unabs $F_{\rm x}^{\rm b}$}                                            & \colhead{$L_{\rm x}^{\rm b,c}$}                   & \colhead{$\chi^2/\nu$}\\
\colhead{} & \colhead{($10^{20}$ cm$^{-2}$)} & \colhead{(keV)}     & \colhead{(solar)}     & \colhead{($10^{-13}$ erg s$^{-1}$ cm$^{-2}$)} & \colhead{($10^{43}$ erg s$^{-1}$)} &  }
\startdata
\hline   \\                  
\swift/XRT       & 1.52    & $4.2^{+1.7}_{-1.4}$ & 0.41 & $1.9^{+0.3}_{-0.4}$ & $2.8^{+0.4}_{-0.7}$ & 54/41	  \\
\chan/ACIS-S     & 1.52    & $2.9^{+1.0}_{-0.7}$ & 0.41 & $1.6^{+0.3}_{-0.2}$  & $2.3^{+0.4}_{-0.3}$ & 230/231 \\
\enddata
\tablecomments{$^{\rm a}$ $90\%$ confidence errors, $^{\rm b}$ Fluxes and luminosities are reported for the $0.5-10$ keV band, $^{\rm c}$ Calculated assuming a luminosity distance of 1101.5 Mpc.} 
\end{deluxetable*}

\section{Weak Lensing results for \cluster} \label{sec:lensing}

To constrain the distribution and total amount of mass in \cluster\ and to search for massive concentrations of dark matter in the other short GRB fields discussed in \S~\ref{sec:WLotherGRB}, we measured the weak shear caused by gravitational lensing. Such measurements were made both from the ground-based {\it VLT}  images and, in the case of \cluster, from the deep {\it HST}/ACS image. Our methodology, described below, follows commonly used procedures for weak lensing measurements. For \cluster, the results are presented in maps of the reconstructed surface mass-density distribution, based on the {\it VLT} and ACS data, and in the form of estimates of the integrated cluster mass. By combining these measurements of the (dark matter dominated) total cluster mass with the X-ray measurements of the hot intracluster gas component discussed in \S~\ref{sec:Xrayanalysis}, a clear physical interpretation of this system emerges, indicating that this is a cluster merger with obvious similarities to the Bullet cluster.

\subsection{VLT data} 
\label{sec:VLTdata}  
 
The weak gravitational lensing measurements were performed using standard IMCAT tools\footnote{http://www.ifa.hawaii.edu/~kaiser/imcat/man/imcat.html}, based on the methodology introduced by \citet{Kaiser1995} and \citet{Luppino1997}, which has been extensively tested against simulated lensing data \citep{Heymans2006}. The weak lensing measurements were made independently on each of the combined images listed in Table~\ref{tab:VLTdata}, and the results for different epochs were only combined in the end, as described below in \S~\ref{sec:results}. For all of these images, the sky background was first set to zero by subtracting a highly smoothed image of the background level as a function of position. Object detection was performed using the IMCAT peak finder ({\it hfindpeaks}), followed by estimation of the local sky background level and gradient around each object ({\it getsky}), aperture photometry ({\it apphot}) and object shape estimation ({\it getshapes}). The latter returned the ellipticity parameters $e_{\alpha}$, calculated from the weighted quadrupole moments of each object  \citep{Kaiser1995}.

Any anisotropies in the point-spread function (PSF) will cause additive errors in the gravitational shear measurements, and such effects must be estimated and removed before the weak lensing effect is measured. For seeing-limited images, this correction is normally well approximated by the expression
\begin{equation}
e_{\alpha}^{\rm cor} = e_{\alpha}^{\rm obs} - P_{\alpha
\beta}^{\rm sm} p_{\beta}
\, ,
\label{eq:PSFcorr1}
\end{equation}
where the smear polarizability $P^{\rm sm}$ given by \citet{Hoekstra1998} is measured by IMCAT ({\it getshapes}), and $p$ is a measure of the PSF anisotropy.  The latter quantity was estimated from $35-700$ non-saturated stellar objects in each {\it VLT } field (the number of available stars depended mostly on Galactic latitude). A second-order polynomial was fit to the PSF anisotropy variation across the field, and the $e_{\alpha}$ values were corrected for PSF anisotropy according to equation~(\ref{eq:PSFcorr1}) (using the IMCAT routines {\it efit} and {\it ecorrect}, respectively).

The isotropic part of the PSF causes a multiplicative error in the shear measurements, which can be corrected using the pre-seeing polarizability tensor $P^{\gamma}$ introduced by \citet{Luppino1997}:
\begin{equation} 
e_{\alpha}^{\rm cor} = e_{\alpha}^{\rm s} + P_{\alpha \beta}^{\gamma} \gamma_{\beta}, 
\label{eq:PSFcorr2}
\end{equation} 
where $\gamma$ is the weak gravitational shear and $P^{\gamma}$ is given by  
\begin{equation} 
P_{\alpha \beta}^{\gamma} = P_{\alpha \beta}^{\rm sh} - P_{\alpha \mu}^{\rm sm} (P_{\alpha \mu}^{\rm sm \star})^{-1}_{\mu \delta} P_{\delta \beta}^{\rm sh \star}.  
\label{eq:Pgamma}
\end{equation} 
Here, $P^{\rm sh}$ is the shear polarizability (from {\it getshapes}) given by \citet{Hoekstra1998}, and asterisks denote the polarizabilities measured for stellar objects.  The quantity $P^{\gamma}$ will vary systematically as a function of the size (e.g., as measured by
the radius $r_g$,
determined by {\it hfindpeaks}) and magnitude of a galaxy, and the $P^{\gamma}$ estimate will be very noisy for individual (small and faint) galaxies. Therefore, in practice these were calculated by binning galaxies in $r_g$-magnitude space (each bin containing typically 50 galaxies), and correcting the ellipticity of each galaxy according to the mean $P^{\gamma}$ value of its bin.  Since the $P^{\gamma}$ tensor is close to diagonal, it is well approximated by a scalar value given by $\frac{1}{2} {\rm Tr}(P^{\gamma})$.  By assuming that the net intrinsic ellipticities of galaxies average to zero, the resulting shear estimator is given by
\begin{equation} 
\hat{\gamma}_{\alpha} = (P^{\rm \gamma})^{-1}_{\alpha \beta} \left[ e_{\beta}^{\rm obs} - P^{\rm sm}_{\beta \mu} p_{\mu} \right].  
\label{eq:gammaest}
\end{equation} 

Finally, a weight was assigned to the $\hat{\gamma}_{\alpha}$ measurement from each galaxy, given by $\langle \gamma^2 \rangle^{-1}$ measured for galaxies in each bin in $r_g$-magnitude space (see above), and these weighted shear estimates were used in the further analysis. The sample of ``background galaxies'' used for the gravitational lensing measurements were selected on the signal to noise ratio of their detection (the $\nu$ value returned by {\it hfindpeaks}), chosen to lie in the range $6 < \nu < 100$. The number of galaxies that passed this selection criterion is listed in the final column of Table~\ref{tab:VLTdata}, and the noise in each mass reconstruction is approximately inversely proportional to the square root of this number.

The projected matter distribution in each field was estimated from the measured shear values, using the maximum probability method introduced by \citet{Squires1996}. We used a regularization parameter of 0.05 and wave modes up to $k=6$ [see \citet{Squires1996} for details].  This produced a $128^2$ pixel image of the dimensionless surface mass density $\kappa = \Sigma / \Sigma_{\rm crit}$, where $\Sigma$ is the physical surface mass density and
\begin{equation} 
\Sigma_{\rm crit} = \frac{c^{2}}{4 \pi\, G} \frac{D_{s}}{D_{l} D_{ls}}, 
\label{eq:sigmacrit}
\end{equation}
where $D_l$, $D_s$ and $D_{ls}$, are angular diameter distances to the lens, to the source, and between the lens and source, respectively. Finally, each of these mass maps was smoothed with a Gaussian of scale $40\arcsec$ to dampen the small-scale noise in the mass reconstruction. The plotted $\kappa$ contour levels in the right panel of figure \label{fig:GRB050509BLightMasscont} start at $\kappa = 0.01$ and are spaced at intervals of $\Delta \kappa = 0.007$. 

\begin{figure*}
\includegraphics[width=1.0\textwidth]{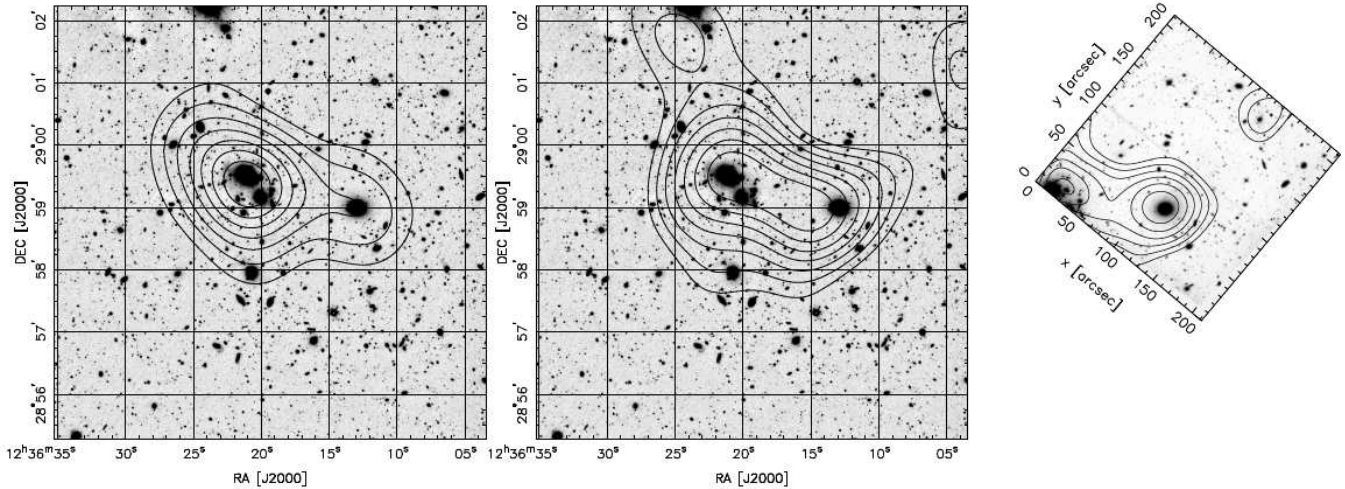}
\caption{The distribution of light (left) and mass surface density $\kappa$ derived from the {\it VLT} data (middle) and from the HST/ACS data (right) in the field of \grb; East is left and North is up. Both distributions from the {\it VLT} data have been smoothed with a Gaussian of scale $40\arcsec$. The light contours in the left panel are derived based on the light distribution of early-type galaxies that form the red sequence in a color-magnitude diagram, and the contour levels are scaled relative to the peak and average value in a similar manner as for the mass map. The contours are superposed on the 3600s $R$-band FORS2 image (see Table ~\ref{tab:VLTdata}). The ACS mass distribution map (right) has been smoothed with a Gaussian of scale 20\arcsec\ and rotated, scaled, and aligned with the {\it VLT} data. The density contours on the right panel are derived based on weak lensing measurements from the {\it HST}/ACS data. The plotted contours start at $\kappa = 0.011$ (which corresponds to $1\sigma$) and are spaced at an interval $\Delta \kappa = 0.011$. The contours are superposed on the 27,480 s combined ACS image.} 
\label{fig:GRB050509BLightMasscont}
\end{figure*}

The significance of features in these mass maps was evaluated by generating randomized catalogues where
the true object positions were kept, and an observed $\hat{\gamma}_{\alpha}$ value was randomly assigned to a given position. Mass maps were then calculated based on these randomized catalogues, and the resulting rms spread in dimensionless surface mass-density, $\kappa$, values provided an estimate of the noise level at each position in the field, thus producing a noise map. A signal-to-noise map of each field was then generated by dividing the $\kappa$ estimate by the noise estimate
(Fig.~\ref{fig:GRB050509BLightMasscont}; middle panel).

\subsection{HST/ACS data} \label{sec:ACSshear}

The weak lensing measurements based on the {\it HST}/ACS data in the field of \grb\ generally followed the methodology described above for the {\it VLT} data, but with some minor deviations, detailed in this section. 

\begin{figure}
\includegraphics[angle=-90,width=0.95\columnwidth]{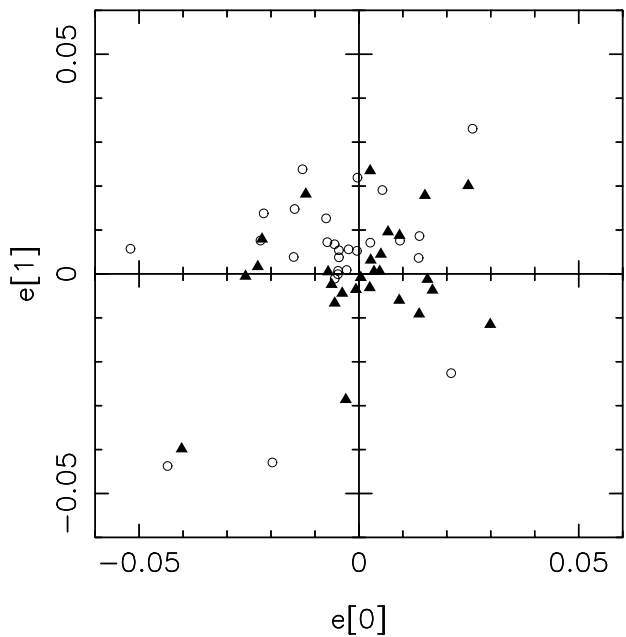}
\caption{Stellar ellipticities in the ACS image of the field of \grb\  prior to (open circles) and after correction for PSF anisotropy (triangles).} 
\label{fig:PSFcorr}
\end{figure}

Objects were detected in the combined F814W image using {\it hfindpeaks}, followed by sky estimation ({\it getsky}), aperture photometry ({\it apphot}) and shape estimation ({\it getshapes}).  Their ellipticities were corrected for the effect of PSF anisotropy by fitting a second-order polynomial to the PSF variation across the field, based on 22 non-saturated stars (see Figure~\ref{fig:PSFcorr} for an illustration of the effect of this correction). Objects in a $\sim 2\arcsec $ wide region at the border of each ACS chip image (where cosmic ray rejection was imperfect and the noise level was elevated, because of dithering) were masked out.  The IMCAT object finder tended to return multiple detections inside some of the brighter galaxies (e.g., from star-forming regions in spiral arms). Such detections were efficiently removed from the final background galaxy catalog, by excluding all sources which had a brighter neighbor within 1\arcsec.

The area covered by the ACS image was 11~arcmin$^2$; $\sim 4400$ objects (detected at $6 < \nu < 250$) were used for the weak lensing analysis, yielding a background galaxy density of  400 objects arcmin$^{-2}$, compared to 40-50 objects arcmin$^{-2}$ for the {\it VLT} data. Hence, weak lensing mass measurements from the {\it HST}/ACS data are approximately $\sqrt{400/45} \simeq 3$ times more sensitive than the ones from {\it VLT}. The background galaxy samples generated from the {\it VLT} and {\it HST} data inevitably have some contamination from cluster galaxies; we describe below in \S~\ref{sec:massest} how this effect was corrected for when making quantitative cluster mass estimates.

We note that for a typical effective background galaxy redshift $z_{\rm bg} \sim 1$, the critical surface mass density given in equation~\ref{eq:sigmacrit} will be close to its minimum value at the redshift of \cluster, making the gravitational lensing effect maximally sensitive to massive structures at this distance.  In addition, a background object surface-density of 400 objects arcmin$^{-2}$ (of which $\sim 270$ objects arcmin$^{-2}$ were detected at $\nu > 10$) is unprecedented for weak gravitational lensing studies of clusters. Within the area covered by the ACS field, our cluster mass reconstruction constitutes the most sensitive weak lensing-based mass measurements thus far for any galaxy cluster.

The higher background galaxy density of the ACS data also enables a mass reconstruction at a finer resolution than from the {\it VLT} data for \cluster. However, the mass distribution near the edges and corners of the ACS field is poorly constrained, given the lack of information about gravitational shear outside the ACS field of view. To overcome this problem, information on the shear external to the ACS field from the wider-field {\it VLT} data was incorporated into our analysis. The combination of HST/ACS-based shear measurements in the smaller area of the cluster core with ground-based data to assess the weak lensing signal over a larger field is a technique used previously by \cite{Clo06} for the Bullet Cluster. Even at the extreme corners of the ACS field, the HST data is the dominant contributor to the derived surface mass density estimate. The smoothing scale of 20\arcsec\ ensures that the contribution of the VLT data to the mass map becomes negligible within the ACS field. 

The projected matter distribution was reconstructed from the measured shear values, using the maximum probability method \citep{Squires1996} with a regularization parameter of 0.05 and wave modes up to $k=12$. This produced a $128^2$ pixel image of the dimensionless surface mass-density, $\kappa$, over the entire field covered by the {\it VLT} data, which was smoothed with a Gaussian of scale $20\arcsec$ (twice the resolution of the {\it VLT}-based mass maps derived in \S~\ref{sec:VLTdata}). The sub-area of this mass map, which overlaps the area observed by ACS, is displayed as contours overlaid on the $F814W$-band ACS image in the right panel of Figure~\ref{fig:GRB050509BLightMasscont}. The ACS image only covers about a quarter of the area of the FORS2 field, and the ACS field is rotated about $133^{\circ}$ counterclockwise with respect to the principal North-South direction, such that
the easternmost peak in the mass and light distribution is located at the corner of the field; the right panel of Figure~\ref{fig:GRB050509BLightMasscont} shows the ACS data aligned to the VLT image. The peak SN ratio value and the rms noise of this mass map is $6.0\sigma$ and $\kappa_{\rm rms}=0.011$, respectively, where the $\kappa_{\rm rms}$ value was calculated in a similar way as for the {\it VLT} data, as described in \S~\ref{sec:VLTdata}.

\begin{figure*}[ht]
\includegraphics[width=0.95\textwidth]{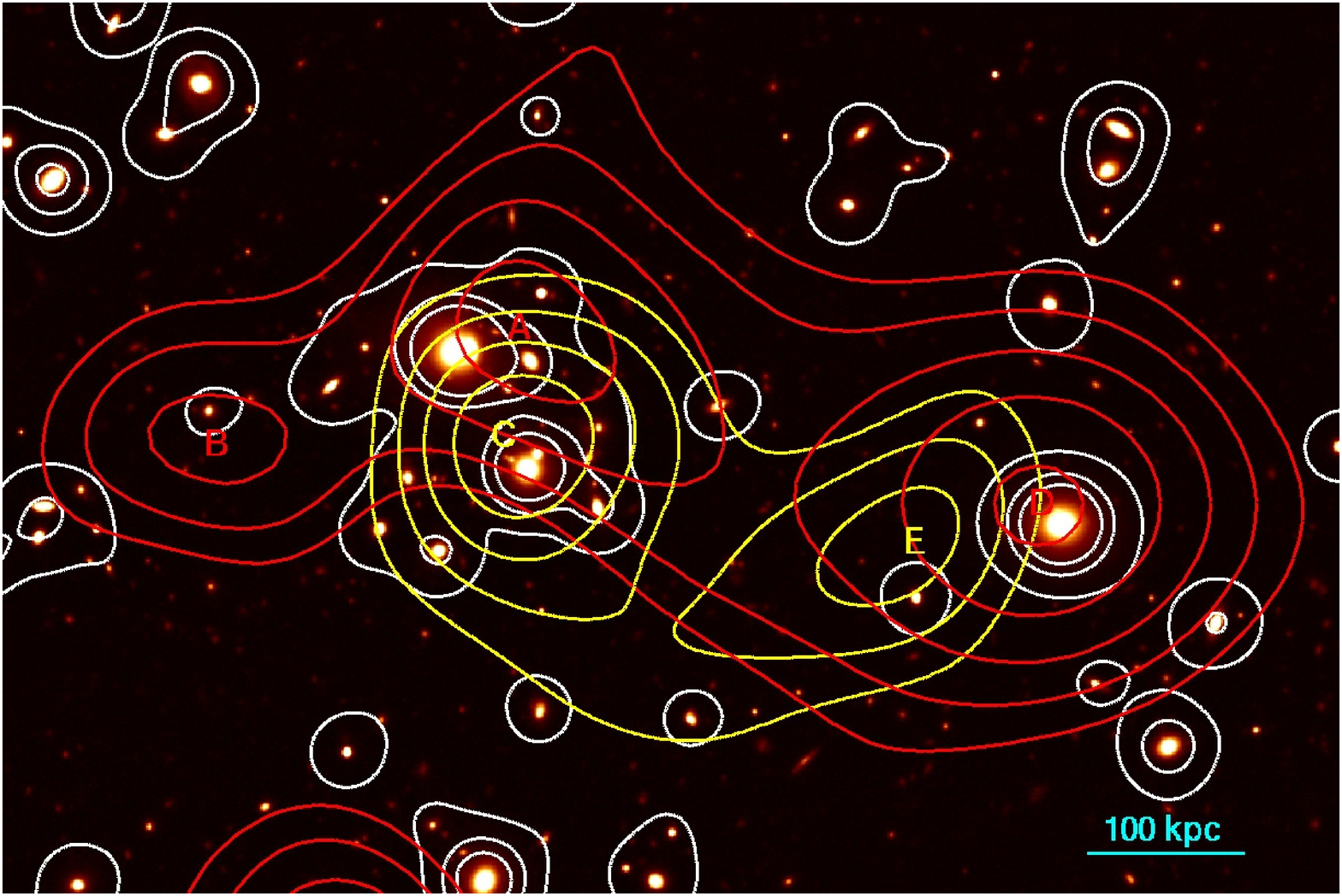}
\caption{The distribution of light (white), mass surface density $\kappa$ (red) and gas (yellow) in the field of \grb; east is left and north is up. The mass density contours are derived based on weak lensing measurements from the {\it VLT} and {\it HST}/ACS data. The light contours are derived based on the light distribution of early-type galaxies that form the red sequence in a color-magnitude diagram, and the contour levels are scaled relative to the peak and average value in a similar manner as for the mass map. The gas (X-ray) contours are from the wavelet transform image filtered with a $20\arcsec$ Mexican hat. The contours are superposed on the 3600s $R$-band FORS2 image (see Table~\ref{tab:VLTdata}). The letters A through E correspond to the peaks of these distributions as described in the text.} 
\label{centroid_combination}
\end{figure*}

\subsection{Cluster Mass Estimate}
\label{sec:massest}

An estimate of the cluster mass can be derived from gravitational lensing measurements, e.g., by fitting the observed one-dimensional shear profile $g_T\, (r)$  to the predicted shear as a function of cluster-centric radius for a given spherically symmetric theoretical mass model, such as an NFW-type mass density profile \citep{Navarro1997}. With two clearly separated components in its mass, light, and X-ray luminous gas distribution, \cluster\ is poorly represented by a single, monolithic structure, at least on scales smaller than or equal to the sub-cluster separation.  However, by using only shear measurements at large radii that encompass both mass concentrations, a fair estimate of the total cluster mass can be made.  

The galaxy shape distortions caused by gravitational lensing provide information on the reduced tangential shear, $g_T = \gamma_T / (1 - \kappa )$, where $\gamma_T$ is the tangential component of the shear. The reduced  $g_T$ was measured in a set of non-overlapping annuli, each with a mean radius $r$ centered on the optical cluster center (i.e., the highest peak in the contours displayed in the left panel of Figure~\ref{fig:GRB050509BLightMasscont}). For the NFW model, the three-dimensional mass-density has a radial dependence of the form, $\rho (r) \propto \left[ (r/r_s)(1+r/r_s)^2 \right]^{-1}$, specified by a scale radius $r_s$ and a concentration parameter $c_{\rm vir} = r_{\rm vir}/r_s$, where $r_{\rm vir}$ is the virial radius of the cluster. N-body simulations of the dark matter distribution in clusters in a $\Lambda$CDM universe give predictions for the dependency of $c_{\rm vir}$ on cluster mass and redshift \citep{Bullock2001,Dolag2004}. The mass of \cluster\ predicted from the X-ray determined mass - temperature relation of \citet{Arnaud2005} is $M_{500c} \simeq 2\times 10^{14} M_{\sun}$, where $M_{500c}$ denotes the mass enclosed within a three dimensional volume bounded by $r_{500c}$. Here, $r_{500c}$ is the radius within which the mean cluster density is 500 times the critical density of the Universe at the redshift of the cluster. Assuming the radial mass-density distribution of the NFW mass model,  this mass translates into $M_{180c} \simeq 3\times 10^{14} M_{\sun}$, for which \citet{Bullock2001} predict a median halo concentration of $c_{\rm vir} = r_{\rm vir} / r_s = 1.14 r_{180c} / r_s = 6.8 / (1+z)$ in a concordance model universe (here, $M_{180c}$ and  $r_{180c}$ are defined in a similar manner as $M_{500c}$ and  $r_{500c}$).  When making the fit, we fix the concentration parameter at this value.  The lensing properties of the NFW model have been calculated by \citet{Bartelmann1996} and \citet{Wright2000}.

As noted above, the shear measurements used for the NFW fit were only made at large cluster-centric radii ($133\arcsec < r < 210\arcsec$), encompassing both sub-clusters of \cluster. Only the {\it VLT} data were used for this fit, as they provided full azimuthal coverage at all radii, unlike the smaller-field ACS data.  Our photometric data in two pass-bands did not allow reliable discrimination between cluster galaxies and lensed background galaxies except for the small fraction of galaxies with observed $V-R$ colors redder than early-type cluster galaxies in \cluster. The former must be at higher redshifts than the cluster galaxies to attain such red $V-R$ values for any normal galaxy spectral energy distribution. The faint galaxy catalogs which were used to measure the gravitational shear are thus significantly contaminated by cluster galaxies that dilute the lensing signal by an amount given by their local sky density. To correct for this contamination, a radially dependent correction factor was applied to the shear, based on the estimation of the magnitude of this effect for a sample of clusters of similar richnesses and redshifts by \citet{pedersen07}.

The distances to the background galaxies, entering into equation~(\ref{eq:sigmacrit}), need to be known to convert the observed lensing signal into a cluster mass estimate in physical units. The background galaxy redshifts were estimated from spectroscopic and photometric redshifts in the Hubble Deep Field [for details, see \citet{Dahle2002}]. The average value of the ratio between the lens-source and observer-source angular diameter distances, $\beta \equiv D_{ls}/D_s$, was calculated given the weights assigned to shear measurement from each galaxy; we estimated a mean $\langle \beta \rangle = 0.694$ for the {\it VLT} data of \cluster .

Our NFW fit yields a mass $M_{180c} = (4.08 \pm 1.56) \times 10^{14} M_{\sun}$ and $r_{180} = 1.44 \pm 0.20$\,Mpc, or $M_{500c} = (2.81 \pm 1.07) \times 10^{14} M_{\sun}$ and $r_{500} = 0.91 \pm 0.13$\,Mpc. The latter radius corresponds to 4\farcm 2 on the sky, i.e., the $6\farcm 9 \times 6\farcm 9$ FORS2 field virtually covers the entire cluster volume enclosed by $r_{500}$. Our cluster mass estimate is consistent with the value $M_{500c} = 2.2^{+3.3}_{-0.6} \times 10^{14} M_{\sun}$ derived by \citet{Pedersen2005}, based on their analysis of the \swift/XRT data combined with the X-ray determined mass - temperature relation of \citet{Arnaud2005}.  Our mass is also consistent with the value $M_{500c} = (2.01 \pm 0.67) \times 10^{14} M_{\sun}$ derived from the \chan/ACIS-S temperature measurement given in Table~\ref{globalspec}, using the X-ray derived mass-temperature relationship of \citet{Vik+06} (which is very similar to the \cite{Arnaud2005} relationship). A somewhat higher, but still consistent, value  $M_{500c} = (3.14 \pm 1.42) \times 10^{14} M_{\sun}$ is estimated using the \chan\ temperature and the weak-lensing based mass-temperature relationship derived by \citet{pedersen07} (assuming a fixed slope $\alpha = 1.5$). Finally, our lensing-derived mass estimate implies that \cluster\ has about 25\% the mass of the Bullet cluster \citep{Clo06,Bra06}.

We also made individual mass estimates of the two subclusters. This was done by measuring the projected 2-dimensional masses using the ``aperture densitometry'' statistic of \cite{Fahlman94}. This statistic measures the mean projected density within an aperture, minus the mean density in a surrounding annulus, which provides a lower bound on the mass contained within the aperture. At the larger radii employed for our NFW model fit described above, it is impossible to unambiguously disentangle the relative mass contributions of the Western and Eastern subclusters. Hence, we choose an aperture of $28\arcsec$ (=100\, kpc at the redshift of the cluster) and measured the mean reduced tangential shear within two annuli defined by ($28\arcsec < r < 75\arcsec$); the first was centered on the \grb\ host galaxy and the second on the global maximum of the cluster light distribution, associated with the Eastern subcluster (see Figure~\ref{fig:GRB050509BLightMasscont}). The outer radius of $75\arcsec$ was chosen to be smaller than the $\simeq 100\arcsec$ separation between the two clusters, in order to distinguish the mass of each subcluster. For the Western subcluster containing \grb\ we derive a mass estimate $M_{\rm ap} (< 100\, {\rm kpc}) = (1.27 \pm 0.26) \times 10^{13} M_{\sun}$, from weak lensing measurements based on the ACS data. The Eastern subcluster is not adequately covered by the superior ACS data, and we therefore used the VLT data to similarly derive a mass $M_{\rm ap} (< 100\, {\rm kpc}) = (2.06 \pm 1.27) \times 10^{13} M_{\sun}$ for this component. Although the uncertainty of the VLT-based mass estimate of the Eastern subclump is fairly large, these values are consistent with the picture provided by the observed distributions of X-ray luminosity and stellar mass (see the following section), which both indicate that the Eastern subcluster is the more massive of the two subclusters.

\subsection{\grb\ and Host Cluster Results} \label{sec:results}  

We discuss below the possible association of \grb\ and \cluster. The \swift/XRT error circle for \grb\ was centered in the outskirts of a giant elliptical galaxy (\galaxy) within the cluster \cluster\  ($z=0.2214$). As detailed above, we were able to recover a significant weak gravitational lensing signal from this cluster, using both ground- and space-based imaging data.

All the mass maps derived for this field, based on the four different combined {\it VLT} images (see Table~\ref{tab:VLTdata}), show evidence for a binary structure in the mass distribution, with one mass peak close to the optical light center of the cluster, and the other close to the giant elliptical galaxy which is the apparent host of \grb. The peak levels in these mass maps correspond to $\sim 4\sigma$. To further improve the accuracy of the mass reconstruction, the four different mass maps were averaged, and the resulting mass contours are plotted in the middle panel of Figure~\ref{fig:GRB050509BLightMasscont}. The left panel in the Figure shows the distribution of light from early-type galaxies inside the cluster, smoothed on the same scale as the mass distribution. This ``light map'' was derived by identifying the tight red sequence formed by early-type cluster galaxies in a $V-R$ vs. $R$ color-magnitude diagram, and selecting all galaxies within an interval of width $\Delta (V-R) = 0.25$, centered on the red sequence. The light map shows a binary structure similar to the mass map with the Eastern peak more dominant, although the total masses of the two subclusters are similar as we discuss below. This indicates that the mass-to-light ratio is somewhat higher for the Western peak,  which coincides with the location of \grb.  While the galaxy light from the Eastern peak is dominated by two early-type galaxies (2MASX J12362010$+$2859080 and 2MASX J12362094$+$2859290), the light of the Western peak is dominated by the apparent host of \grb. 

Based on the 2MASS $JHK_s$ photometry of these three galaxies, we estimate the stellar mass of the two sub-peaks as $(2.5 \pm 1.2) \times 10^{11} M_{\sun}$ and  $(5.2 \pm 1.9) \times 10^{11} M_{\sun}$ for the Western and Eastern peaks, respectively. These estimates were based on modeling of the spectral energy distribution, using the SED models of elliptical galaxies from \cite{Sil+98} for stellar populations ranging in age from 1.5 Gyr to 13 Gyr. The systematic and random errors of the stellar mass determinations from such SED modeling are estimated to be approximately a factor of 2 \citep{Bel+07,Mic+08,Mic+10}.

The mass map recovered from the ACS data (see Figure~\ref{fig:GRB050509BLightMasscont}, right panel) shows a significant mass peak ($6\sigma$) centered on the \grb\ host galaxy, which coincides spatially with the position of \grb\ (within the respective errors of the \grb\ position and the mass peak centroid). This mass peak has an elongation at a position angle similar to the \grb\ host galaxy, pointing towards the easternmost peak. Together, the weak lensing mass reconstructions from the ACS data in Figure~\ref{fig:GRB050509BLightMasscont} (right panel) and from the {\it VLT} data in Figure~\ref{fig:GRB050509BLightMasscont} (middle panel) reveal a consistent picture of two sub-clumps of similar mass, separated by $\simeq 100\arcsec$. The morphology and position of the two mass peaks (particularly the Western peak, which is located well within the ACS field) are most firmly constrained by the ACS data, given the significantly higher background galaxy density in the space-based images. We discuss the light, mass and gas peak separations in detail in section \S~\ref{sec:discuss}. 
 
\section{Weak Lensing Survey of Other Short GRBs}
\label{sec:WLotherGRB}

The serendipitous discovery of a merger in the field of \grb\ motivated a search in the public data of other short GRBs for similar mergers. Establishing their presence in more than one GRB field would prove extremely significant for our understanding of the origin of the short GRB phenomenon. We selected, therefore, all public {\it VLT} data (up to the end of this study) taken during follow-up observations of well-localized short GRBs, resulting in five additional event fields. Using the same weak lensing technique described above (\S \ref{sec:VLTdata}), we estimated the mass concentration along their lines of sight in the archival {\it VLT} images.  The results of our analysis are described below. 

\subsection{GRB~050724}

The host galaxy of this burst has a measured redshift of $z=0.258$ \citep{Prochaska2006}. Spectroscopy of other field galaxies reveal an apparent galaxy overdensity at $z=0.30$, but no clustering of galaxies was seen in redshift space at the burst redshift \citep{Berger2007}. The mass map of this field does not show any significant peaks at $> 2\sigma$. The field is at low Galactic latitude, with a Galactic extinction $E(B-V) = 0.59$, estimated from the dust maps of \citet{Schlegel1998}. This produces $\sim 2$ magnitudes of extinction in the  $V$-band and $1.1$ magnitudes of extinction in the $I$-band. In addition, the field is heavily crowded by foreground stars, which further reduces the number of useful background galaxies.  Hence, a very massive ($M \sim 10^{15} M_{\sun}$) GRB host cluster would be needed to produce a detectable lensing signal in this field. The absence of such a signal is consistent with the non-detection of a host cluster by \cite{Berger2007}.

\subsection{GRB~050813}

The mass reconstruction of this field does not reveal any significant peaks at $> 2\sigma$. The field was observed during poor conditions (clouds), and the high background sky level results in a somewhat reduced number of detected galaxies (see Table~\ref{tab:VLTdata}), and, consequently, a higher noise level in the mass reconstruction. 

\subsection{GRB~050906}

Two mass peaks are visible in this field, a northern peak, visible at the $\sim 3\sigma$ level, and  a southern peak at the $2-3 \sigma$ level. These peaks are at a level where they could easily be caused by random fluctuations in the orientations and ellipticities of the background galaxies, and hence, we are unable to report a firm detection of any mass concentration in this field.  

\subsection{GRB~050911}

This GRB has a rather large error circle of radius $2\farcm 8$ \citep{Page2006} which coincides with the known cluster EDCC 493. \citet{Berger2007} measure a redshift $z \sim 0.1646$ and velocity dispersion $\sigma_v = 660^{+135}_{-95}\, {\rm km s}^{-1}$ for this cluster. They also estimate a mass of $M_{500c} \simeq 2.5\times 10^{13} M_{\sun}$ for EDCC 493, based on an estimate of the X-ray temperature from diffuse emission seen in the \swift/XRT data, combined with the mass-temperature relation of \citet{Arnaud2005}. This is an order of magnitude lower than the estimated mass of \cluster, the host cluster of \grb, suggesting that the mass of EDCC 493 would be too small for a firm detection via weak lensing. 

We did recover one peak in the mass map of this GRB field at the $\sim 3.5\sigma$ level; the rough structure of the mass map bears some similarity to the distribution of brighter galaxies in this field. However, the highest mass peak we recovered is offset from the peak of the X-ray emission (which is spatially coincident with the brightest cluster galaxy of EDCC 493) by $\sim 3\arcmin$. In addition, this mass peak is located in a region of fairly high projected galaxy density, and the redshifts of the brightest galaxies in this region show a wide scatter \citep{Berger2007}, suggesting that the mass peak, if real may simply be a projection effect of unrelated structures over a wide range of redshifts. A second, lower peak (at $\sim 2.5 \sigma$) is seen closer to the position of the X-ray and optical center of EDCC 493. The lensing detection of mass is thus at best at a marginal level in this field, and so we cannot firmly conclude that we have detected any mass overdensity associated with  GRB~050911.  It is, however, worth noting that this is the GRB field with the second most significant mass detection after \grb\ and also the only other field known to contain a galaxy cluster.

\subsection{GRB~060313}

The mass map of this field does not reveal any significant peaks at $> 2\sigma$.  The lensing analysis was based on a combined image from all available FORS1 and FORS 2 $R-$band data with a total exposure time of almost 5 hours.

\section{Discussion}
\label{sec:discuss}

Here, we discuss some implications from our X-ray and optical observations of the cluster \cluster.
In particular, we consider the implications of the association of this cluster and its galaxy environment for models of the origin of short GRBs. We also discuss the dynamical state of the cluster, and suggest that, with further observations, it might be a useful object to help constrain the properties of dark matter.

\subsection{\grb\ - \galaxy\ association}

The association between \grb\ and \galaxy\ is difficult to understand if the GRB resulted from any mechanism involving massive stars and recent star formation \citep{Lee07}, such as a collapsar \citep{Woosley1993}.  Such an association would instead suggest that there is at least a non-negligible fraction of short GRB progenitors with lifetimes greater than about 7 Gyrs, which can outburst within a Hubble time. Among early type galaxies approximately 50\% of the stellar mass content is in galaxies with $M > 10^{11} M_\sun$ that typically reside in clusters (Zheng \& Ramirez-Ruiz 2007). Since it is likely that galaxies in clusters shut off their star formation process early on, a long progenitor lifetime further supports the tendency for short GRBs to happen in cluster galaxies. Indeed, \galaxy\ shows no indications of UV emission or optical emission lines \citep{Prochaska2005}. The optical spectrum of the galaxy is just that of a typical giant elliptical, with no evidence for star formation within the last several Gyr. 

On the other hand, \galaxy\ is a very propitious site for the merger of neutron stars or any other mechanism involving an old, close binary containing at least one compact object. Although it is not possible to detect radio pulsars at the distances of even nearby giant elliptical galaxies, \chan\ observations have shown that these galaxies have very large populations of low mass X-ray binaries (LMXBs) containing accreting neutron stars and black holes \citep{SIB00,ALM01}. This is particularly true of central dominant cluster galaxies. These systems are the likely progenitors to recycled millisecond pulsars. In giant ellipticals (particularly cDs), a high fraction ($\ga$50\%) of the LMXBs are located in globular clusters \citep[GCs; e.g.,][]{SKI+03}. This presumably occurs because close binary systems containing compact objects can be formed dynamically in GCs \citep{Hut83,Grindlay06,Lee10}. While there is less direct evidence that close double neutron star binaries can form easily in GCs, the double neutron star system PSR B2127+11C in the Galactic GC M15 \citep{AGK+90} is an example of such a system, and has a sufficiently short gravitational radiation lifetime to merge in $\sim 2 \times 10^8$ yr. The large bulge luminosities and very high specific frequencies of GCs in giant ellipticals and particularly cDs can explain their large populations of LMXBs. Thus, cluster-dominant giant elliptical galaxies like \galaxy\ are very good locations for neutron star - neutron star or neutron star - black hole mergers \citep{Zemp2009}.

If \grb\ is due to a neutron star - neutron star merger (or other close binary mechanism) in \galaxy, then there is reasonable probability that it occurred in a GC in this galaxy \citep{Lee10}. GCs have a broader radial distribution in elliptical galaxies than field stars \citep[e.g.,][]{Har91}. Note that \grb\ is probably located in the outer regions of the galaxy, which might be more likely if it came from a binary merger progenitor. LMXBs tend to be associated preferentially with the more luminous GCs in a galaxy \citep{SKI+03}. If we assume an absolute magnitude $M_{\rm I} \ga -11$ for the GC, its apparent magnitude at the distance of \galaxy\ would be $m_{\rm I} \ga 29$. Such a GC might be detectable as a nonvariable, red, point-like optical counterpart to \grb\ in our deep {\it HST} image. However, given the large positional uncertainty for the GRB, it would be very difficult to establish a reliable association at this magnitude level. 

Assuming the association between \grb\ and \galaxy\ is correct, then \grb\ occurred in a much more diffuse gaseous environment than GRBs in star forming regions of galaxies. In general, elliptical galaxies have very low density interstellar gas. Our \chan\ image of \galaxy\ and the cluster \cluster\ (Fig.~\ref{xrayimages}) shows that \grb\ probably occurred in a region dominated by intracluster gas, with a low ambient density likely effecting the strength and nature of the afterglow \citep{pana01}. If the burst occurred instead in a GC, the resulting afterglow could then at least in part be due to the interaction of the relativistic ejecta with the stellar winds of the red giant cluster members. Due to the large stellar density in the star cluster core,  the external shock would then take place within a denser medium than the IGM \citep{Lee10}.

\subsection{Dark Matter in \cluster}

Our  adaptively smoothed \chan\ image (Fig.~\ref{xrayimages}), our wavelet transform images (Fig.~\ref{montage}), and our power ratio results (Fig.~\ref{powers}) all show that the X-ray surface brightness distribution of the cluster \cluster\ is bimodal. The most luminous clump of X-ray emission is centered (Table~\ref{Xcentroids}) near the brightest cluster galaxy, 2MASX J12362094$+$2859290, which is about a half a magnitude brighter than the host of \grb, while the second clump of X-ray emission is centered just slightly to the east of the GRB host galaxy, \galaxy. The distribution of the galaxy light in the cluster shows a similar bimodal structure (left panel in Fig.~\ref{fig:GRB050509BLightMasscont}), again with the Eastern subcluster containing more galaxies, galactic light and stellar mass (see \S~\ref{sec:results}). Our {\it VLT} weak lensing map of the mass surface density (middle panel in Fig.~\ref{fig:GRB050509BLightMasscont}) shows that the mass is also distributed with a bimodal distribution, although it appears to be somewhat more symmetric than the X-ray gas or galaxies. Moreover, our {\it HST}/ACS weak lensing map (Figure~\ref{fig:GRB050509BLightMasscont}, right panel) indicates that the bulk of the mass in the Western subcluster is centered on the GRB host galaxy \galaxy, while the X-ray emission is displaced to the east. 

Fig.~\ref{centroid_combination} shows the X-ray (gas), light and mass contours overplotted on the {\it HST}/ACS field. The X-ray contours of the western subcluster associated with the GRB host galaxy \galaxy\ shows an E-W elongated distribution of emission, with a drop in the X-ray surface brightness to the west of the X-ray peak and to the east of the galaxy. The galaxy itself is in a relatively faint region of cluster X-ray emission. This morphology suggests that the X-ray gas might be in a ``cold front'' \citep{Mar+00,VMM01}. A possible interpretation is that the western X-ray clump was originally a cool core centered on the galaxy \galaxy, and that this galaxy was at the center of one of the two subclusters. In this model, the X-ray gas has been displaced by ram pressure due to the gas in the main subcluster

\begin{deluxetable}{lccr} 
\tablecolumns{4} \tablewidth{0pt} 
\tablecaption{Centroid locations of the cluster main mass substructures \label{Mcentroids}}
\tablehead{ \colhead{Substructure} & \colhead{RA} & \colhead{DEC}  \\
     &    \colhead{(h m s)} & \colhead{(\arcdeg\ \arcmin\ \arcsec)}  }
\startdata
Eastern Cluster	 A$^*$ & 12 36 20.094 & $+$28 59 33.62  \\
Eastern Cluster	 B & 12 36 24.162 & $+$28 59 13.24 \\
Western Cluster D	 & 12 36 13.043 & $+$28 59 02.64  \\
\enddata
\tablecomments{$^*$We follow the same letter designation as shown on Fig.~\ref{centroid_combination}. The estimated uncertainties in the centroid values are $\approx3.3\arcsec$.} 
\end{deluxetable}

This geometry may be similar to the merger in the Bullet cluster, 1E$0657-558$ \citep{Mar+02}, where the X-ray emission shows a cold front and a bow shock associated with the merging subcluster; it is somewhat different than the geometry in MACS J$0025.4+1222$ \citep{Bradac08} where there is only one gas (X-ray) peak between the light and mass peaks. Here and in the Bullet cluster, the X-ray gas is displaced to the east of the subcluster dominant galaxy and other galaxies associated with the subcluster. In all cases, lensing measurements indicate that the total mass and dark matter are centered on the galaxies in the subclusters, and do not follow the X-ray gas centroids. 

We have used the wavelet transform results described in Section~\ref{sec:wavelet} to quantify the relative separation of the X-ray and mass centroids in both subclusters of \cluster. We identified two X-ray and three mass centroids; Table \ref{Mcentroids} lists the coordinates of the mass centroids using the letter designation shown in Fig.~\ref{centroid_combination} and Table~\ref{Xcentroids} gives the coordinates of the X-ray centroids. The separation between the two Eastern mass centroids ($A, B$) and the Eastern X-ray centroid ($C$) is $\delta S = 19\farcs0$ ($68$ kpc) and $50\farcs9$ ($182$ kpc), respectively, while the separation between the Western mass centroid ($D$) and the Western X-ray centroid ($E$) is $\delta S = 23\farcs4$ ($84$ kpc). We note that these separations are much larger than the uncertainties in the X-ray and mass centroids (2\arcsec\ and 3.3\arcsec\ respectively). 

As noted in the case of the Bullet cluster \citep{Clo04,Mar04,Bra06,Clo06,Clo07} and MACS J$0025.4+1222$ \citep{Bradac08} such systems provide a number of very strong tests of the existence and nature of non-baryonic dark matter. More recently, six additional merging clusters have been found with separated dark matter and baryonic components (e.g., Dawson et al. 2012 and references therein). The fact that the surface density of the total mass and of the dark matter aligns with the galaxy distribution is not in accord with the expectations of theories in which dark matter is actually the result of a different theory of gravity, but all of the mass is in baryonic material. In the cluster \cluster\ as in most rich clusters, the overwhelming majority of the baryonic matter is in the hot intracluster gas. Thus, in any theory in which baryons are the primary source of gravity, the gravitational potential and the apparent dark matter distribution should follow the gas in a cluster, and not the galaxies. The lensing observations of all cluster mergers thus far require the presence of non-baryonic dark matter.

The second test of the nature of dark matter comes from the agreement between the location of the centroid of the dark matter and that of the galaxies \citep{Ran08}. The galaxies form a collisionless fluid, which is why they are not subject to the same ram pressure as the X-ray gas. The distribution of the dark matter indicates that it also must be a nearly collisionless fluid. In the Bullet cluster, the implied upper limit for the dark matter self-interaction cross-section \citep{Mar04,Ran08} is considerably smaller than the values required by models in which dwarf galaxy mass distributions are the result of cores in the dark matter due to self-interaction. Unfortunately, we cannot derive the same limits for \cluster\ since our X-ray observations are not deep enough to allow the detailed surface brightness profile and spectra in the possible cold front region of the western subcluster to be determined. Deeper imaging with \chan\ and spectroscopy (with \chan\ or {\it XMM-Newton}) are needed to determine the parameters of this merger, and allow \cluster\ to be used as a test of dark matter.

\section{Conclusions}
\label{sec:conclusion}

Since \grb\ was the first short GRB for which an X-ray afterglow was found, its putative association with \cluster\ motivated multiple follow up observations in all wavelengths. Our \chan\ observations revealed the merger nature of the cluster, which in turn provided the impulse for weak gravitational lensing measurements pointing to the effects of \dm. The X-ray image analysis using the power ratio technique placed \cluster\ among the most irregular of the clusters at the same redshift range and among clusters of $z>0.5$, indicating that the merger is likely very recent (not relaxed) and highly disturbed. The centroids of light and mass in the cluster coincide and are offset from the gas centroids. These offsets range between $70$ and $180$ kpc strongly supporting the collisionless nature of \dm. \cluster\ is now among the less than a dozen of dissociative systems providing a strong test of the nature of dark matter. However, \cluster\ is less massive and fainter in X-rays. A much longer dedicated \chan\ observation is needed to determine the spectral properties of the gas accurately, and derive the merger velocity and kinematics for this system. Since \cluster\ is at a relatively low redshift ($z=0.22$), a deeper \chan\ and/or {\it XMM-Newton} image might also detect a bow shock associated with the merger, as was seen in the Bullet cluster.

\acknowledgments
We thank A. von der Linden for her detailed and critical remarks on the paper. HD acknowledges support from the Research Council of Norway, including a postdoctoral research fellowship. CLS was supported by the National Aeronautics and Space Administration through {\it Chandra} awards GO7$-$8129X, GO8$-$9085X, GO9$-$0135X, GO9$-$0148X, and GO1$-$12169X, and through {\it HST} awards HST$-$GO$-$10582.02$-$A, HST$-$GO$-$10835.01$-$A, HST$-$GO$-$11679.01, HST$-$GO$-$12012.02$-$A, and HST$-$GO$-$12202.0$-$A. Support for LAL was provided by NASA through the Einstein Fellowship Program, grant PF1--120085; LAL also acknowledges support by the Pappalardo Fellowship in Physics at MIT. CK thanks the DARK Cosmology Center, who provided the nourishing environment for the final stages of this paper. AJvdH was supported by an appointment to the NASA Postdoctoral Program at the MSFC, administered by Oak Ridge Associated Universities through a contract with NASA. The Dark Cosmology Centre is funded by the Danish National Research Foundation. The WSRT is operated by ASTRON (Netherlands Institute for Radio Astronomy) with support from the Netherlands foundation for Scientific Research. Some of the results were based on observations made with ESO Telescopes at the La Silla or Paranal Observatories under programmes ID 075.D-0261, 075.D-0415, 075.D-0468, 075.D-0787,076.D-0612.

\nocite{*}
\bibliographystyle{apj}
\bibliography{grbcluster}

\end{document}